\documentclass[10pt,prd,longbibliography,superscriptaddress,floatfix]{revtex4-1}

\usepackage{graphicx}
\usepackage{amsmath}
\usepackage{esvect}
\usepackage{stmaryrd}
\usepackage{amssymb}
\usepackage{systeme}
\usepackage{bbold}
\usepackage{natbib}
\usepackage{breakcites}
\usepackage[margin=1.2in]{geometry}
\usepackage{multirow}

\usepackage{rotating}

\usepackage{url}

\graphicspath{{figures/}}
\newcommand{\nico}[1]{\textcolor{black}{#1}}
\newcommand{\nicob}[1]{\textcolor{black}{#1}}

\usepackage[colorlinks,citecolor=blue]{hyperref}
\begin{document}

\title[]{Dynamical systems for remote validation of very high-resolution ocean models}

\author{
Guillermo Garc\'ia-S\'anchez
}
\address{Instituto de Ciencias Matematicas, CSIC, C/Nicolas Cabrera 15, Campus Cantoblanco, Madrid, 28049, Spain}
\address{Escuela T{\'e}cnica Superior de Ingenieros de Telecomunicaci\'on, Universidad Polit{\'e}cnica de Madrid, Av. Complutense, 30, Madrid,28040,Spain}

\author{
 Ana M. Mancho$^*$
}
\address{Instituto de Ciencias Matematicas, CSIC, C/Nicolas Cabrera 15, Campus Cantoblanco, Madrid, 28049, Spain}

\author{
 Antonio G. Ramos
}
\address{ Instituto ECOAQUA, Faculty of Marine Sciences. Campus Universitario de Tafira, Universidad de Las Palmas de Gran Canaria, 35017 Las Palmas de Gran Canaria, Spain}
\author{
 Josep Coca
}
\address{ Instituto ECOAQUA, Faculty of Marine Sciences. Campus Universitario de Tafira, Universidad de Las Palmas de Gran Canaria, 35017 Las Palmas de Gran Canaria, Spain}
\author{
 Jose Antonio Jimenez Madrid
}
\address{Instituto de Ciencias Matematicas, CSIC, C/Nicolas Cabrera 15, Campus Cantoblanco, Madrid, 28049, Spain}
\date{\today}

\date{\today}

\maketitle
	
\section*{Abstract}

This paper presents and investigates a novel methodology for validating high-resolution ocean models using satellite imagery. \nicob{High-resolution ocean models provide detailed information in coastal areas where other available data products are too coarse. Models are usually fitted by comparing them with observations; However, accessing in situ data in all small coastal areas is not feasible, as in situ observations are scarce and obtained through dedicated ships or instruments in limited and selected regions. Our work aims to use alternative remote sensing information to overcome this challenge.}
The approach involves establishing connections between the satellite observations and the outcomes of various \nicob{ computational experiments carried out using the Regional Ocean Modeling System (ROMS), which allows the selection of different parameters to run the ocean model.} These choices are not fully determined a priori and \nicob{ each one produces distinct outputs, which are then linked to the images through dynamical systems objects}.
By defining a performance index, we are able to quantify which experiment provides a better representation of the local ocean state.


\section{Introduction} 

Ocean numerical models help to understand and predict ocean variables that have an impact on the environment, climate, and human activities. 
Oceanic data produced by models are increasingly becoming available, with various services providing valuable information on the ocean's physical state and dynamics. For instance, the Copernicus Marine Environment Monitoring Service (CMEMS) offers regular and systematic data on current and future variables, as well as retrospective data records (re-analysis) of the global and European regional seas. Alternatively, the HYbrid Coordinate Ocean Model (HYCOM) --a multi-institutional effort sponsored by the National Ocean Partnership Program (NOPP) as part of the U.S. Global Ocean Data Assimilation Experiment (GODAE)-- is dedicated to developing and evaluating data-assimilative hybrid coordinate ocean models.
The products offered by these services typically have resolutions of the order of 2.5-8 km and have been  successful in describing events above those scales \cite{mh370, oleg, GGS2022}. More recently  it has been pointed out \cite{GGS2021} the necessity of using products with much higher resolution to address  management needs at the coastal level. These needs may include 
 navigation, harmful algae blooms, effluent dispersion, search and rescue operations, oil spill response, etc. In these cases, models are particularly valuable for understanding and predicting ocean currents  for the management of aspects that affect  life in these areas, including beach conditions and near-shore environments.

In situ measurements  are incorporated into ocean models through different methodologies to achieve outputs that are accurate and consistent with the real ocean state. Among the instruments employed for in situ measurements are moored current meters, drifters, shipboard current profilers that  measure sea surface velocity, etc.
The deployment and maintenance of this equipment is expensive and, in general, its spatial coverage is limited to hotspot areas where the investment in these measurements is worth it. 
There exist however wide coastal and marine protected areas, where high-resolution models can be very valuable for management purposes,  but for which in situ measurements are not available and therefore their validation is challenging. 
However, observations in these areas are possible through
satellites that today are equipped with multiple remote sensing systems capable of providing valuable information on the dynamic properties of extensive oceanic coastal areas. In this work, we will explore the
use of methodologies based on dynamical systems and satellite imagery for feedback and adjustment
of ocean model outputs and as a validation method. 




More specifically,  this study starts with the
creation of a very high-resolution hydrodynamic model in an area for which no in situ measurements were available. This is   the area of the Rafina port in Greece, which was taken as a study case in the framework of the \href{http://www.impressive-project.eu/}{IMPRESSIVE} project in which authors participated. Sentinel 2 and Sentinel 3 images near the coast provide  images in the optical band that after being processed are able to identify  turbidity caused by suspended particles, including particulate and dissolved organic and inorganic matter. These materials may  derive from sources such as primary production blooms and wastewater. The spread of these compounds makes visible  convoluted shapes, which are linked to  invariant dynamical objects, called Lagrangian Coherent Structures, obtained from the currents.  Thus, patterns visible in remote sensing images can be correlated to patterns obtained from the hydrodynamic model. We will utilize this idea to explore an effective methodology for validating the transport capacity of the model outputs. This approach differs from classical validations that compare in situ measurements of currents, sea level, etc. with those obtained from the models.
Our comparison is systematically implemented by adjusting the parameters and parameterizations required by the model, which include closure relationships for unresolved turbulent scales, lateral boundary conditions for the nested model, and the coupling of surface ocean motion with atmospheric winds, among others. In addition, an index is defined to quantify the degree of fitting of each parameter set to the satellite observations.
   
The manuscript is structured as follows. Section \ref{hm} provides a description of the equations and software utilized to set up the hydrodynamical model for the Rafina Port. It also discusses the boundary conditions and settings specific to the temporal window of October-December 2019. In Section \ref{rs}, the available remote sensing products for that temporal window are described. Section \ref{ds} delves into the dynamical systems concepts used to establish the connection between the hydrodynamical model outputs and the satellite imagery. Section \ref{nexppq} outlines the various settings for the experiments conducted and details the procedure used to quantify their performance. The results are explained in Section \ref{res}, and finally, \nicob{ a discussion and the conclusions are presented } in Section \ref{con}.
\section{The hydrodynamic model}
\label{hm}

\subsection{The oceanic regional model}
To set up the hydrodynamic model we have used the Regional Oceanic Modeling System
(ROMS) \nico{(see details on \cite{romsmanual})}. 
\nico{ROMS numerically solves the Navier–Stokes equations while considering the hydrostatic and Boussinesq assumptions \cite{Shchepetkin2005, Haidvogel2000}. However, ROMS does not solve these 3D equations exactly in a turbulent regime; instead, it utilizes the unsteady Reynolds-Averaged Navier–Stokes (RANS) equations. These are 
derived from the Navier-Stokes equations} by applying Reynolds decomposition, which decomposes the instantaneous velocity field ($\textbf{v}_T$) into a time-averaged ($\overline{\textbf{v}}_T$) and a fluctuating component ($\textbf{v}'_T$).
 In this work, the mean velocity is the velocity field solved in equations \eqref{ns1}-\eqref{ns3}, which is referred to as $\textbf{v} = (u, v, w)$ there.  The fluctuating component $\textbf{v}'_T= (u', v', w')$ is solved using turbulent closure \nico{relations, which are detailed later in this section}.
ROMS has been employed in numerous operational oceanography systems on regional scales.
Examples include the \href{https://tidesandcurrents.noaa.gov/models.html}{NOAA operational forecast systems} for locations such as Tampa Bay, Chesapeake Bay, and the Gulf of Maine, and the SAMOA system, which the Spanish Public State Port Agency developed to offer tailored operational meteorological and oceanographic data \cite{SAMOA}. ROMS has also been utilized in diverse applications like estuarine modeling \cite{cerralbo2015hydrodynamic}, sediment transport \cite{grifoll2014formation}, oceanic basin modeling \cite{malanotte2000water}, climate research \cite{di2008north}, and the study of atmosphere-wave-current interactions \cite{warner2010development}. A key advantage of ROMS is its modular structure and efficient parallelization using the Message Passing Interface (MPI), enabling rapid computational performance \cite{wang2005parallel}.

ROMS solves the governing dynamical equations in flux form. The vertical \nico{takes} terrain-following coordinates that are named sigma vertical coordinates. We have used Cartesian coordinates in the horizontal, although other choices are available.  These equations take the form \cite{Haidvogel2008}: 
\begin{equation}\label{ns1}
    \frac{\partial u}{\partial t} + \textbf{v} \cdot \nabla u - f v = - \frac{\partial \phi}{\partial x} -  \frac{\partial}{\partial z}\left(\overline{u'w'} - \nu \frac{\partial u}{\partial z} \right) + \mathcal{F}_u + \mathcal{D}_u, 
\end{equation}

\begin{equation}\label{ns2}
    \frac{\partial v}{\partial t} + \textbf{v} \cdot \nabla v + f u = - \frac{\partial \phi}{\partial y} -  \frac{\partial}{\partial z}\left(\overline{v'w'} - \nu \frac{\partial v}{\partial z} \right) + \mathcal{F}_v + \mathcal{D}_v.
\end{equation}
In the Boussinesq approximation, changes in density are disregarded in the momentum equations, except for their influence on the buoyancy force within the vertical momentum equation. When the hydrostatic approximation is employed, it is additionally presumed that the buoyancy force is counterbalanced by the vertical pressure gradient:
\begin{equation}\label{ns3}
    \frac{\partial \phi}{\partial z} = -\frac{\rho g}{\rho_0},
\end{equation}
where $\phi = p/\rho_0$ and $p \approx -\rho_0 g z$ is the total pressure. The continuity equation for an incompressible fluid:
\begin{equation}\label{cont}
    \frac{\partial u}{\partial x} +\frac{\partial v}{\partial y} +\frac{\partial  \omega }{\partial z}  = 0.
\end{equation}
Here $\textbf{v}=(u,v, \omega)$ and $u$, $v$, and $\omega$ are the components of velocity in the horizontal ($x$ and $y$) and vertical (scaled sigma coordinate, $s$) directions respectively; and $f$ is the Coriolis parameter. An over-bar represents a time average, and a prime ($'$) represents turbulent fluctuations. $\rho$ and $\rho_0$ are total and reference densities; $g$ is the acceleration due to gravity; $\nu$ is the  molecular viscosity. Here all $\mathcal{F}$ are forcing terms and $\mathcal{D}$ are horizontal diffusive terms.


\begin{table} 
\centering
\begin{tabular}{|c|c|c|c|c|c|c|c|c|c|c|c|c|c|c|c|c|}
\hline                                     
Parameters& $p$ & $m$  & $n$ & $\sigma_k$&   $\sigma_\psi$  &$c_\mu^0$& $c_1$ &$c_2$&$c_3^+$&$c_3^-$&$k_{\textrm{min}}$&$\psi_{\textrm{min}}$&$F_{\textrm{wall}}$&$c$& $\nu$&$\nu_{\theta}$\\ \hline
Values for & & & & & & & & & & & & & & & &\\ 
$k-\varepsilon$ model & 3  & 1.5  & -1  & 1.0  & 1.3   & 0.5544 & 1.44  & 1.92  & 1.0 & -0.518  & $7.6\cdot 10^{-6}$  & $1.0 \cdot 10^{-12}$ & 1.0   & 1 & $5.0 \cdot 10^{-6}$ & $5.0 \cdot 10^{-6}$                             \\ \hline
\end{tabular} 
\caption{Parameters for the generic length scale turbulence- $k-\varepsilon$ model and Kantha-Clayson stability functions   (see \cite{warner2005performance}).} \label{tab:parametroske}
\end{table}

Temperature ($T$) and salinity ($S$) scalar tracers, generically denoted as $C$, are coupled to the momentum equations,  and their transport equation is given by:
\begin{equation}\label{treq}
   \frac{\partial C}{\partial t} +\textbf{v} \cdot \nabla C  =  - \frac{\partial}{\partial z}\left(\overline{c'w'} - \nu_\theta \frac{\partial C}{\partial z} \right) + C_{\textrm{source}} + \mathcal{F}_C + \mathcal{D}_C.
\end{equation}
Here $C_{\textrm{source}}$ represents scalar source/sink terms; $\nu_\theta$ is the molecular diffusivity.
Again $\mathcal{F}$ are forcing terms and $\mathcal{D}$ are horizontal diffusive terms.
A function $ \rho = \rho(T, S, p) $ is required to specify the equation of state. We have selected the option of the nonlinear equation of state introduced by \citep{nonlinearEOS}. 

The horizontal viscous and diffusivity terms $\mathcal{D}$  are given by the harmonic expressions:
\begin{align}
\mathcal{D}_X &= A_X\nabla^2 X, \quad \textrm{for} \ X = \{u,v,T,S\},
\end{align}
where $A_{u,v}$ is the horizontal viscosity coefficient for $u, v$ and $A_{T}$ and $A_S$ are the horizontal diffusivity coefficients for $T, S$. 
In our application, we will explore optimal settings of $A_{X}$ for $X=\{u,v,T,S\}$ according to values given in Table \ref{tab:experiments}.

\nicob{Eqs.} \eqref{ns1}-\eqref{ns2} and \eqref{treq}  are closed by parameterizing the Reynolds stresses and turbulent tracer fluxes as:
\begin{equation}\label{cl}
    \overline{u'w'} = -K_M \frac{\partial u}{\partial z}; \quad \overline{v'w'} = -K_M \frac{\partial v}{\partial z}; \quad \overline{c'w'} = -K_H \frac{\partial \rho}{\partial z},
\end{equation}
with
\begin{equation}
K_M= c\sqrt{2k}l S_M+\nu, \, \, \, \, \, \,\, \, \, \, \, \,  K_H=c\sqrt{2k}l S_H+\nu_{\theta}. \label{kmkh}
\end{equation}
In these equations, $K_M$ represents the eddy viscosity for momentum and $K_H$ represents the eddy diffusivity for tracers. $S_M$ and $S_H$  are stability functions that describe the effects of shear and stratification. From the several possibilities available in ROMS,   we have chosen the Kantha-Clayson stability functions (see \cite{warner2005performance}). The replacement of Eqs. \eqref{cl} into  Eqs.\eqref{ns1}-\eqref{ns2} and \eqref{treq} leads to the standard harmonic form for vertical viscous/diffusive terms. The turbulent quantities $k$ (turbulence kinetic energy) and $l$ (turbulent length scale) must
be determined in order to close the set of equations.
ROMS contains a variety of methods to do this. 
In this work, we  considered the generic lengthscale (GLS),  proposed by \cite{umlauf2003generic} and implemented in ROMS by \cite{warner2005performance}. This  is a generic two-equation turbulence closure scheme. The first equation represents the conventional transport equation for the turbulent kinetic energy ($k$), 
\begin{equation}
    \frac{\partial k}{\partial t} + {\bf v} \cdot \nabla k  = \frac{\partial}{\partial z}\left( \frac{K_M}{\sigma_k}\frac{\partial k}{\partial z} \right) + P +B - \varepsilon,
\end{equation}
where $\sigma_k$ is the turbulence Schmidt number for $k$.
$P$ and $B$ represent production sources by shear and buoyancy, and $\varepsilon$ is the dissipation. They are given by:
\begin{equation}
P  = K_M \left[\left(\frac{\partial u}{\partial z}\right)^2 + \left(\frac{\partial v}{\partial z}\right)^2\right], \,\,\,\, B =  K_H \left( \frac{g}{\rho_0}\frac{\partial \rho}{\partial z}\right), \,\,\,\, \varepsilon = (c^0_\mu)^{3+p/n} k^{3/2+m/n}\psi^{-1/n}, \label{param1}
\end{equation}
here the stability coefficient $c^0_\mu$ is determined based on experimental data for the unstratified channel flow with a log-layer solution. The second equation of  the GLS model describes the evolution of $\psi$,  a generic parameter used to establish the turbulence length scale:
\begin{equation}
    \frac{\partial \psi}{\partial t} + {\bf v} \cdot \nabla \psi = \frac{\partial}{\partial z}\left( \frac{K_M}{\sigma_\psi}\frac{\partial \psi}{\partial z} \right) + \frac{\psi}{k}(c_1P+c_3B-c_2\varepsilon F_{\textrm{wall}}). \label{param2}
\end{equation}
The parameter $\sigma_\psi$ represents the turbulence Schmidt number for $\psi$, indicating the ratio of momentum diffusivity to the diffusivity of $\psi$. 
Coefficients $c_1$ and $c_2$ are chosen to 
be consistent with experimental observations of decaying homogeneous, isotropic turbulence \cite{wilcox1998turbulence}. The parameter $c_3$ assumes different values for stable ($c_3^+$) or unstable ($c_3^-$) stratification. Also,
\begin{equation}
    \psi = (c^0_\mu)^p k^m l^n. \label{turbf}
\end{equation}
The parameters in expressions \eqref{ns1}-\eqref{turbf} for the $k-\varepsilon$ model, which is considered in our simulations, take values summarized in Table \ref{tab:parametroske} (see \cite{warner2005performance}). \nicob{This choice has been utilized in \cite{SAMOA} for settings similar to those considered in this work.}


We further explain the sigma vertical coordinates used in ROMS. The variables related to this coordinate are $\zeta$, representing the wave-averaged free-surface elevation, and $h$, denoting the depth of the sea floor relative to the mean sea level. In a topography-following vertical coordinate system, there is a transformation, 
\begin{equation}\label{eq:z}
    z = z(x,y,\sigma),
\end{equation}
where $z$ is the Cartesian height and $\sigma$ is the vertical distance from the surface measured as the fraction of the local water column thickness; i.e, $-1\leq \sigma \leq 0$, where $\sigma = 0$ correspond to the free surface, $z =\zeta$, and $\sigma = -1$ corresponds to the oceanic bottom or bathymetry, $z = h(x,y)$. In the case of the classical $\sigma$-coordinate, (\ref{eq:z}) reduces to

\begin{equation} \label{eq:z2}
    z = \sigma \cdot h(x,y).
\end{equation}

This may be combined with nonlinear stretching, $S(\sigma)$,
\begin{equation}\label{eq:z3}
    z(x,y,\sigma) =  S(\sigma) \cdot h(x,y).
\end{equation}


ROMS allows to use a general vertical coordinate system $S$-coordinate \cite{Song1994}:

\begin{equation}
    z(x,y,\sigma) = \zeta(x,y,t) + \left[ \zeta(x,y,t) + h(x,y)\right] S(x,y,\sigma),
\end{equation}
\begin{equation}
    S(x,y,\sigma) = \frac{h_c\sigma + h(x,y)C(\sigma)}{h_c + h(x,y)},
\end{equation}
where $h_c \leq h_{min}$ is a free parameter that must be adjusted for each application.
$h_c$ is a depth above which the vertical grid spacing of the
sigma layers becomes (a) nearly uniform and (b) nearly independent
of local depth, $h$, as long as $h \gg h_c$. In our case  $h_c$ = 84 m
and $C(\sigma)$ can take different forms. In our application, we are assuming that 
\begin{equation}
    S(x,y,\sigma) = 
     \begin{cases}
      0 \quad \textrm{if} \ \sigma = 0, \ C(\sigma) = 0, \ \textrm{at the free-surface}, \\
      -1 \quad \textrm{if} \ \sigma = -1, \ C(\sigma) = -1, \ \textrm{at the ocean-bottom}, 
    \end{cases}\,
\end{equation}
and $C(\sigma)$ is defined as a continuous, double-stretching function controlled at the surface  by $\theta_s$ and at the bottom layer by $\theta_b$. In our particular case, we have selected $\theta_S = 5$ and $\theta_B = 0.4$. The refinement function is then defined in two steps. First, we consider for the surface:
\begin{equation}
    C_s(\sigma) = \frac{1 - \cosh(\theta_S \sigma)}{ \cosh(\theta_S) - 1}, \quad \textrm{for} \ \theta_S > 1.
\end{equation}
Then we consider the bottom refinement function:
\begin{equation}
    C(\sigma) = \frac{\exp(\theta_B C_s(\sigma)) - 1}{ 1 - \exp(-\theta_B)}, \quad \textrm{for} \ \theta_B > 0.
\end{equation}

\subsection{The geographical domain }
\begin{figure}
	\centering
	a)\includegraphics[width=0.4\linewidth]{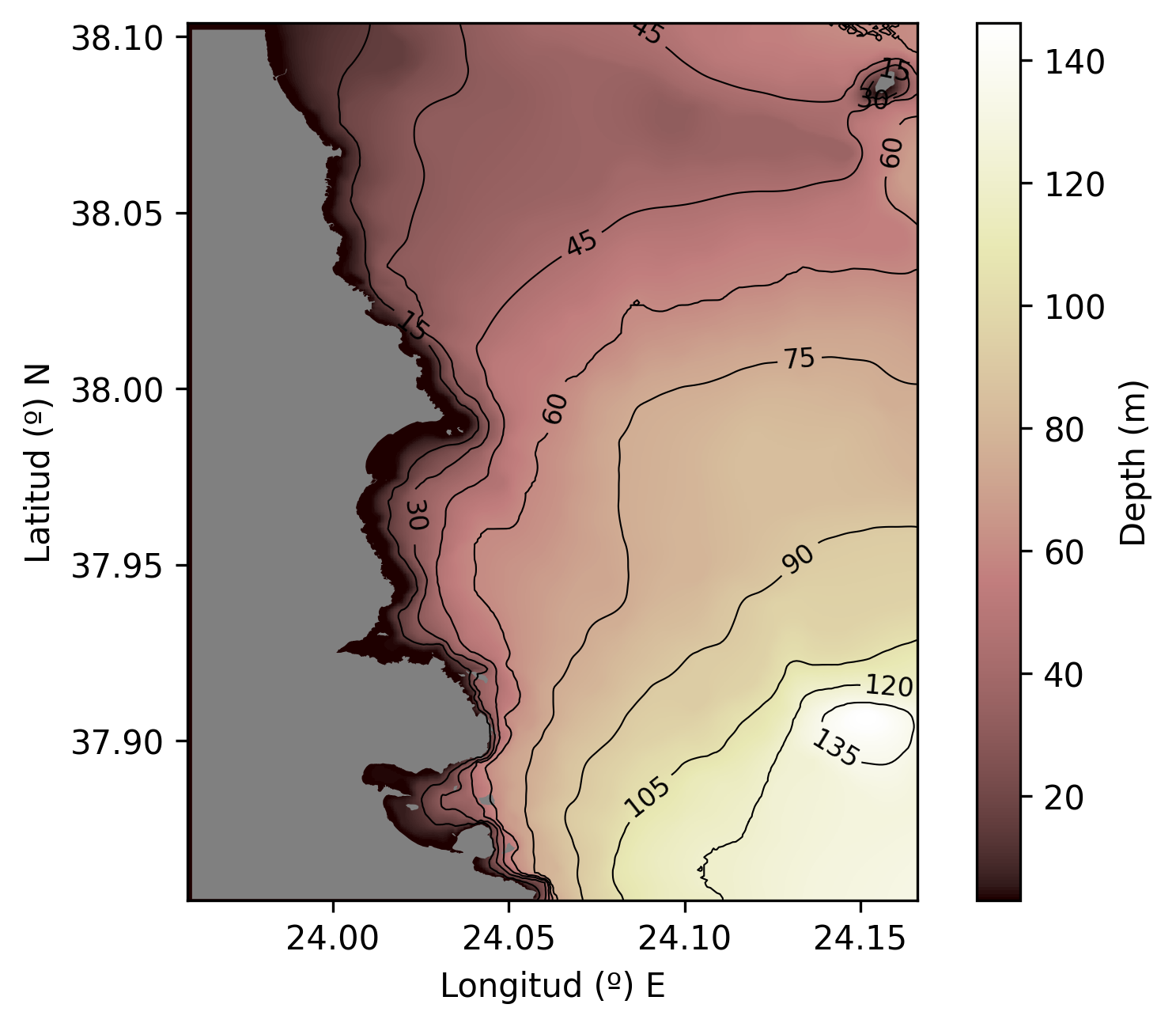} b)\includegraphics[width=0.4\linewidth]{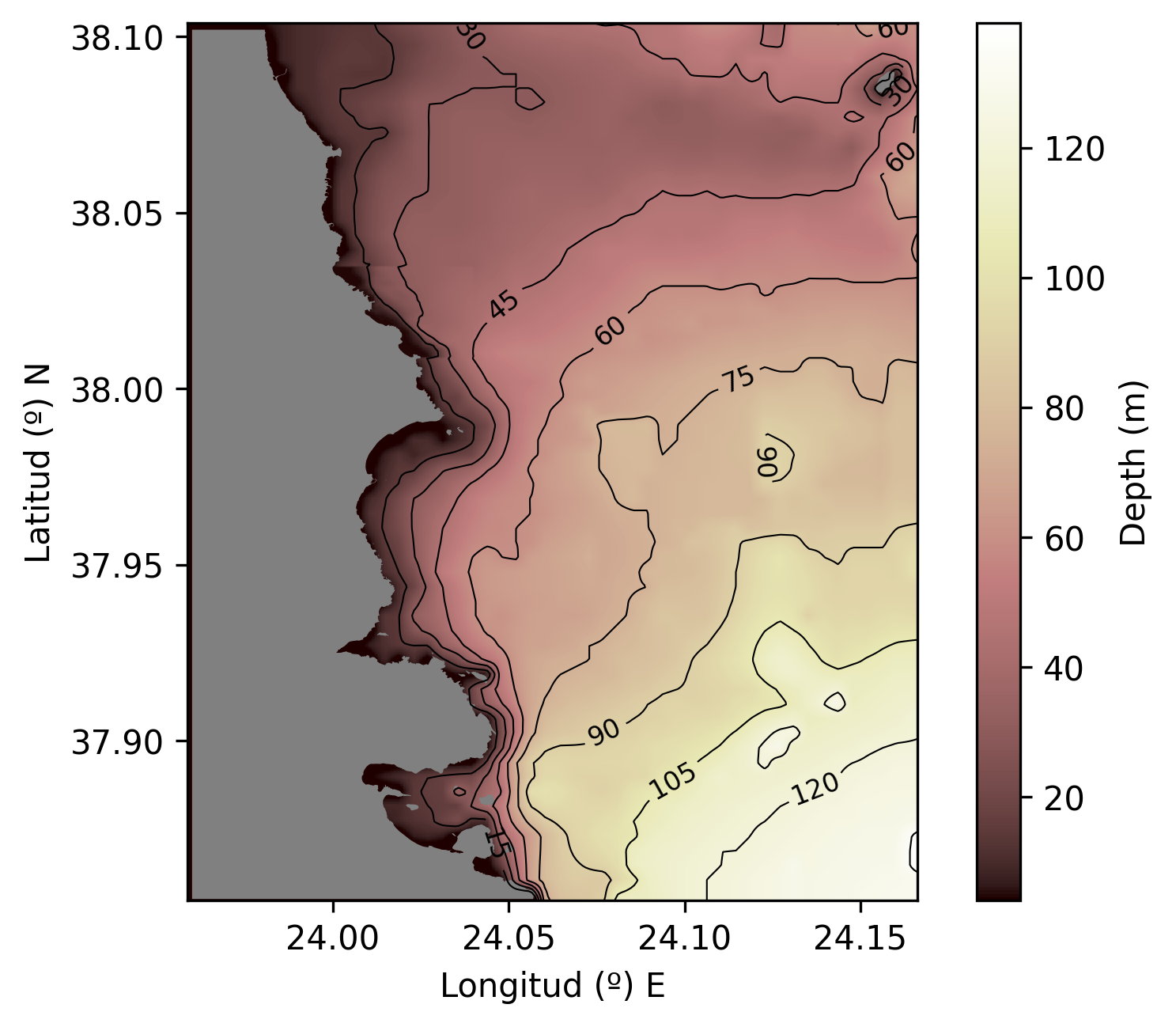}
	\caption{ Bathymetry of the Rafina area in the domain $L_0$. The color intensity
corresponds to the depth of the sea floor, with brighter colors indicating greater depths. a) NAVIONICS source data; b) GEBCO source data. 
}
\label{fig:rafina}
\end{figure}
Our study sets up a hydrodynamic model in the neighborhood of Rafina Port. The area lies on the Aegean Sea coast, east of the Penteli mountains and northeast of the Mesogaia plain. It is 5 km (3 mi) north of Artemida, 7 km (4 mi) south of Nea Makri, and 25 km (16 mi) east of Athens city center.
Specifically, the selected domain $L_0$ is between latitudes 37.8541-38.1042 degrees North and longitudes 23.9583-24.1667 degrees  East. 

Figure \ref{fig:rafina} illustrates the domain where equations are solved with two different choices for the function $h(x,y)$ representing the sea bed.
Panel a) shows a bathymetry in the area obtained from contour levels that follow a NAVIONICS chart, which we have computed with a  resolution of up to $10\times 10$ meters. Panel b)  displays the sea floor offered by \href{https://www.gebco.net/}{GEBCO}. This is a publicly-available bathymetry of the world's oceans. It operates under the joint auspices of the International Hydrographic Organization (IHO) and the Intergovernmental Oceanographic Commission (IOC) (of UNESCO). Its resolution is about 15 arc seconds. 
Sigma-coordinates employed by   ROMS require that topographic steepness is limited to prevent pressure gradient errors. For this reason, topographies in Figure \ref{fig:rafina} are smoothed by selectively applying a local filter to reduce the r-factor 
below 0.2. The r-factor is a measure of the sharpness of the variations in the bathymetry. It is defined as the ratio of the maximum height difference between adjacent points ($\Delta h$) to the average height of those points ($2h$); see, for example, \cite{HaidvogelBook}. Additionally, all depths in $L_0$ shallower than 0.5 m are reset to 0.5 m.

Although the appearance of both bathymetries is very similar we will explore their influence on the model outputs in section \ref{res}. 
Indeed, the sea floor has a number of impacts on ocean circulation. It helps steer large-scale circulation, and small gaps in the ocean floor can even influence the direction of dominant currents. In addition, small-scale bathymetric features can also affect ocean circulation. When ocean currents flow over the rough sea floor, energy is converted from horizontal flow into vertically propagating waves, leading to increased vertical mixing over rough topography \cite{Gille2003}.

\subsection{Discretization and numerical algorithms }

Regarding the spatial resolution implemented in the domain of interest, we have  adopted a horizontal resolution of 80$\times$ 80 meters, which is achieved  with the  grid of dimensions $Lm \times Mm$ reported in Table \ref{table:L0}. The horizontal discretization implemented by ROMS is on 
an Arakawa `C' grid and spatial derivatives are computed with
a centered, second-order
finite difference approximation.

\begin{figure}
	\centering
 \includegraphics[width=1\linewidth]{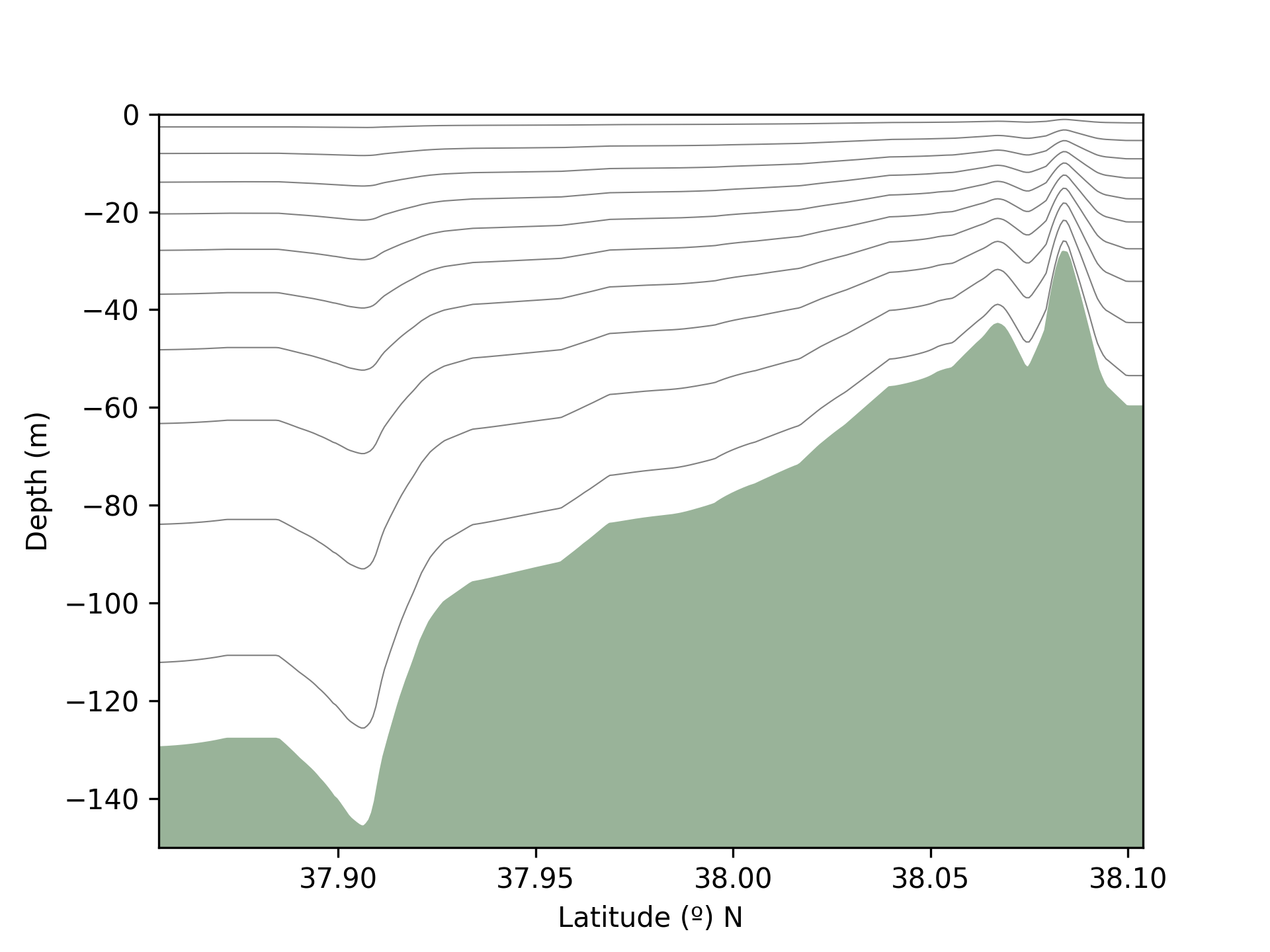}	
	\caption{Vertical discretization at longitude 24.15 $^\circ$E. The green color represents the bathymetry, while the lines indicate the different $\sigma-$levels based on the stretching function utilized. }
\label{fig:vertical_disc}
\end{figure}

Discretization of the vertical coordinate introduces a set of coordinate surfaces,
\begin{equation}
    \{  z_{k+\frac{1}{2}} = z_{k+\frac{1}{2}}(x,y), k= 0,1,\cdots N \}.
\end{equation}
In particular, for our application we have considered $N=10$, that is  10 sigma levels. Figure \ref{fig:vertical_disc} illustrates the 
 $\sigma-$levels used at longitude 24.15 $^\circ$E.
 
The stability of 
numerical models are affected by the numerical time-integration method, the spatial discretization, and the permitted modes in the equations. \nicob{
The temporal integration is conducted with a decomposition of the 3D fields into a barotropic part (obtained by averaging the 3D equations in depth) and a baroclinic part. First, a time step is chosen using a $CFL$ criterion.} Advection is characterized by a velocity scale, $U$. The shortest advective timescale in a numerical model is $\Delta/U$, where $\Delta$ is the spatial grid-scale and $U$ is a measure of the maximum characteristic velocity.
The $CFL$ number is the ratio of this characteristic timescale to the model time step: 
\begin{equation}
    CFL = \frac{U\Delta t}{\Delta}.
\end{equation}
In the case of ROMS' time scheme, $CFL$ must be less than a number on the order of unity. Then, the largest time step that can support numerical stability for a given flow and grid spacing scales is
\begin{equation}
    \Delta t^u_{\textrm{max}} \propto \frac{\Delta}{U}.
\end{equation}
As we said before, $\Delta = 80$ meters so we consider $\Delta t = 10$ seconds. \nicob{In what follows, this time step is referred to as $\Delta_{t}$.}


ROMS uses a computational kernel that combines temporal averaging of the barotropic mode with modified pressure-gradient terms to accurately resolve barotropic processes and reduce errors in the density field, while maintaining the efficiency of the split-explicit formulation. In this process, two time steps are used to advance 3D momentum and tracer variables in the numerical model to effectively capture the complex interactions between the barotropic and baroclinic modes in the ocean. The first time step ($\Delta_{t}$) is used to compute the right side of the 3D momentum equations, while the second time step ($\Delta_{tt}$) is used for updating the barotropic mode. \nicob{Table \ref{table:L0} provides the values considered for these time steps in our simulations.} Vertical interpolation is carried out using either fourth-order centered schemes or conservative parabolic splines. The combination of moderate-order spatial approximations, quasi-monotone advection operators, and enhanced conservation properties leads to more robust, accurate, and less diffusive solutions than those produced by earlier terrain-following ocean models \cite{Haidvogel2008}. More details on the numerical methods used in ROMS can be found in the literature \cite{Shchepetkin2005, Shchepetkin1998, Shchepetkin2003, romsmanual}.

\subsection{Boundary Conditions}
The ocean near coastlines experiences dynamic interactions at various scales. Coastal currents are influenced by local wind forcing, tides, and remote factors transmitted offshore and along the coastal waveguide. Offshore currents may be generated as part of the large-scale circulation or emerge from near-shore regions. However, it is impractical to accurately resolve all types of currents. Instead, regional models can be used to calculate local currents under the influence of local forcing and the larger-scale circulation, given a suitable prescription for lateral boundary forcing \cite{Marchesiello2001}. We detail next how boundary conditions are chosen in order to implement an appropriate downscaled high-resolution model.

\subsubsection{Bottom boundary layer (BBL)}
The dynamics of the bottom boundary layer (BBL) affect the stress on the flow exerted by the bottom, which is incorporated into the Reynolds-averaged Navier–Stokes equations as boundary conditions for momentum. ROMS can use either of two sub-models to represent BBL processes: (i) simple drag-coefficient expressions, or (ii) more complex formulations that consider the interactions of waves and currents over a moveable bed. In our application, we use the former.

\begin{equation}
    K_M \frac{\partial u}{\partial z} = \tau^x_b(x,y, t); \quad  K_M \frac{\partial v}{\partial z} = \tau^y_b(x,y, t); \quad  K_H \frac{\partial T}{\partial z} = 0; \quad K_H \frac{\partial S}{\partial z} = 0.
\end{equation}

The drag-coefficient methods implement formulae for linear bottom friction, quadratic bottom friction, or a logarithmic profile. For our application, we have considered the quadratic bottom friction:
\begin{equation}
    \tau^x_b(x,y, t) = C_d^b \cdot u \sqrt{u^2+v^2}; \quad   \tau^y_b(x,y, t) = C_d^b \cdot v \sqrt{u^2+v^2},
\end{equation}
\begin{equation}
    \tau^x_b(x,y, t) = C_d^b\cdot (u-u_b) \|\textbf{u} - \textbf{u}_b\|, \quad   \tau^y_b(x,y, t) = C_d^s \cdot (v-v_b) \|\textbf{u} - \textbf{u}_b\|,
\end{equation}
where $u_b, v_b$ refers to the eastward and northward bottom speed components and ${\bf u}=(u, v)$. We consider that $u_b, v_b = 0$. The values we will explore of $C_d^b$ are given in Tables \ref{tab:bs} and \ref{tab:experiments}.

We also examined the effect of the oceanic topography's bottom drag in two different scenarios: free-slip and no-slip. In the no-slip scenario, the normal and tangential components of the fluid velocity field are both equal to zero at the interface between a moving fluid and a stationary wall. In the free-slip scenario, the normal component of the fluid velocity field is equal to zero at the interface, but the tangential component is unrestricted. This condition is also known as the no-penetration condition.

\subsubsection{Atmospheric forcing}
The surface conditions, evaluated at $z=\zeta$ are:
\begin{equation}
    K_M \frac{\partial u}{\partial z} = \tau^x_s(x,y, t); \quad  K_M \frac{\partial v}{\partial z} = \tau^y_s(x,y, t) ; \quad K_H \frac{\partial T}{\partial z} = 0; \quad
    K_H \frac{\partial S}{\partial z} = 0.
\end{equation}
Where $\tau^x_s$ and $\tau^y_s$ are the components of the wind stress acting on the free surface in the $x$ and $y$ directions, respectively. We have assumed there is no exchange of heat and salinity with the atmosphere. Moreover, we have chosen a simple wind stress given by


\begin{equation}
    \tau^x_s(x,y, t) = C_d^s\cdot (u-u_w) \|\textbf{u} - \textbf{u}_w\|, \quad   \tau^y_s(x,y, t) = C_d^s \cdot (v-v_w) \|\textbf{u} - \textbf{u}_w\|,
\end{equation}
where $u_w, v_w$ refers to the eastward and northward wind speed components and ${\bf u}=(u, v)$.
The $C_d^s$ values we have explored  are  given in Tables \ref{tab:bs} and \ref{tab:experiments}.

\begin{figure}
	\centering
 \includegraphics[width=1\linewidth]{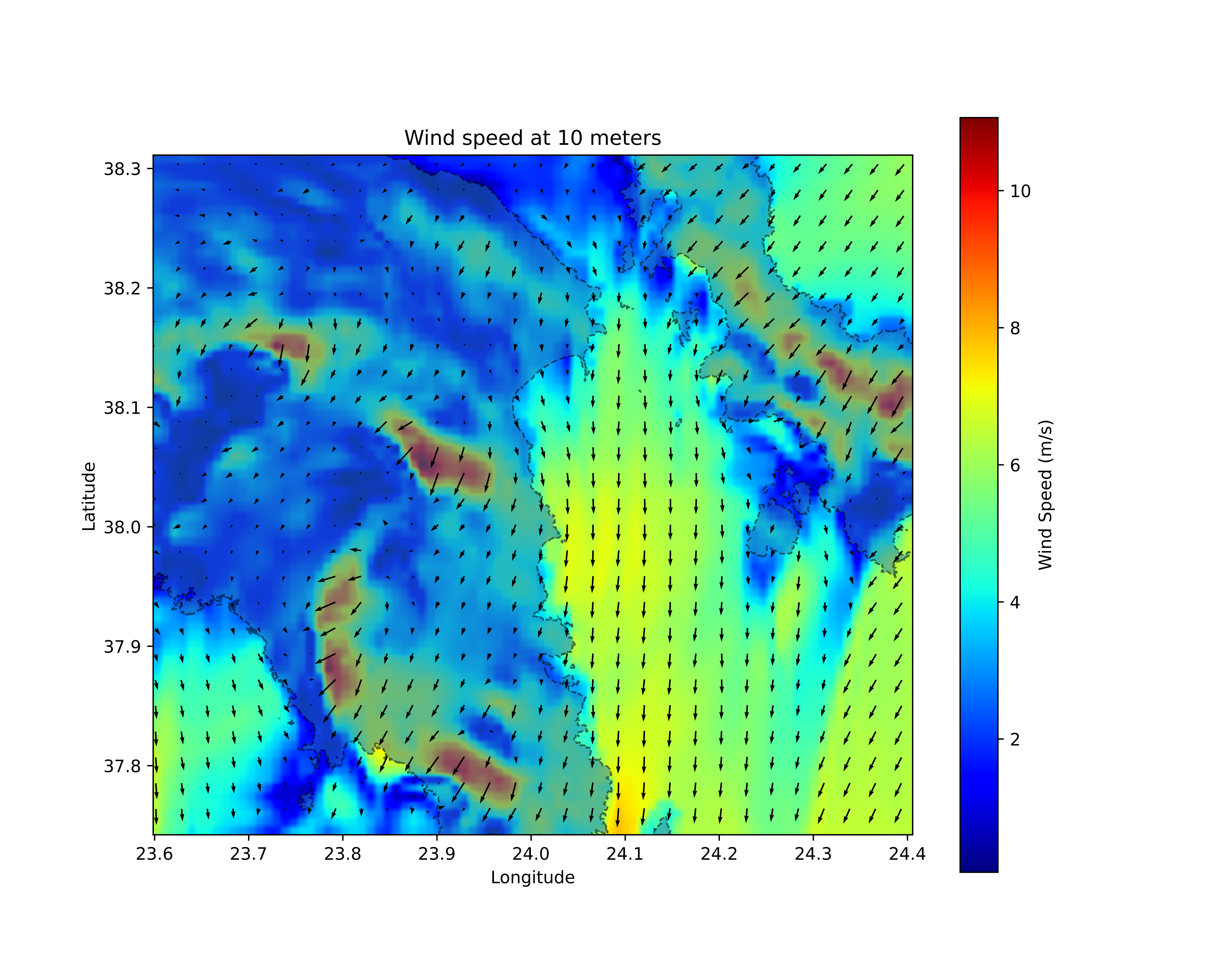}	
	\caption{The figure depicts the WRF domain, with the wind speed intensity at 10 meters above the ground visualized in the background. The visualization corresponds to the specific date and time of November 2, 2019, at 12:00:00. Arrows are employed to indicate the direction of the wind.}
\label{fig:wrf_speed}
\end{figure}

The winds were computed using the NCAR Advanced Research Weather Research Forecasting (ARW) model version 4.0. WRF is a fully compressible, Eulerian non-hydrostatic equations model (with a run-time hydrostatic option) that uses Arakawa-C grid staggering for horizontal discretization and the time-split 3rd ordered Runge Kutta integration scheme. It includes various parameterizations and schemes for microphysical processes, cumulus parameterization, land-surface modeling, boundary layer parameterization, long-wave and short-wave radiation interactions in the atmosphere. 
For this study, the parameterizations and schemes used were as follows: surface layer based on the Monin-Obukhov Similarity scheme, incorporating a Carlson-Boland viscous sub-layer and standard similarity functions \cite{beljaars1995parametrization}; land-surface parametrization from the Thermal Diffusion scheme, using a soil temperature-only scheme with five layers; boundary-layer YSU scheme \cite{hong2006new}; and the Noah land-surface model.
The model domain encompasses the geographical region of the Greek territory, ranging from 23.64 $^\circ$N to 24.36 $^\circ$N in latitude and from 33.7 
 $^\circ$E to 34.35 $^\circ$E in longitude. The horizontal grid spacing is set at 800 meters, and there are 34 vertical levels within the model. The model assimilates ERA5 data, obtained from both single and pressure levels, as a data source \cite{ERA5}. Figure \ref{fig:wrf_speed} displays the plotted wind speed at a height of 10 meters.

\subsubsection{Open boundary conditions}

 In this study, we have set up a model that aspires to provide realistic and informative high-resolution fields in the sense of operational oceanography. This requires  careful nesting on the boundaries and in our setting, we have considered that the CMEMS product in the Mediterranean Sea offers good estimations for the required external fields on the described boundaries below. The Mediterranean Forecasting System's physical component (Med-Physics) consists of a combined hydrodynamic-wave model that covers the entire Mediterranean Basin, including tides. The model has a horizontal grid resolution of about 4 km (1/24$^\circ$) and features 141 unevenly spaced vertical levels. The hydrodynamics are derived from the Nucleus for European Modelling of the Ocean (NEMO v3.6) and incorporate tidal representation, while the wave component comes from Wave Watch-III. Model solutions are adjusted using a 3DVAR assimilation scheme (OceanVar) that incorporates temperature and salinity vertical profiles, as well as satellite Sea Level Anomaly observations along the track \cite{clementi2019mediterranean}.

We define \nico{next} conditions at the lateral boundaries. First, we take into consideration tides that constrain sea surface height and barotropic velocities at the boundaries. \nico{ Tides may be adjusted in ROMS through} the Chapman and Flather boundary conditions, which we detail next.

\textbf{Chapman boundary condition}. This condition for surface elevation was investigated by Chapman \cite{chapman1985numerical}. \nico{It assumes that at an open boundary $\zeta$, the sea surface elevation, satisfies the wave equation: } 
\begin{equation}
    \frac{\partial \zeta}{\partial t}  = \pm \sqrt{gh}\frac{\partial \zeta}{\partial x}, 
\end{equation}
\nico{where $\sqrt{gh}$  is the shallow-water wave speed.}

\textbf{Flather boundary condition.}
\nico{For the normal component of the barotropic velocity, $\overline{u}_n$, it is used the radiation condition proposed by \cite{flather1976tidal}:}
\begin{equation}
    \overline{u}_n = \overline{u}_n^{\textrm{ext}} - \sqrt{\frac{g}{h}} \left(\zeta - \zeta^{\textrm{ext}} \right),
\end{equation}
where $\overline{u}_{\textrm{ext}}$ represents the external barotropic velocity and $h$ is the local water depth. \nico{In this equation, the differences between the external data and the model predictions are permitted to propagate beyond the domain at the speed of the external gravity waves.}

The exterior values are used to provide tidal boundary conditions to the barotropic mode. In limited area coastal domains -as it is our case-  the tides can be treated as being entirely remotely forced, i.e., the influence of the gravitational tide generating forces within the domain is negligible. The principal challenge to modeling tides accurately is the proper representation of tidal elevation and velocity boundary conditions around the entire model perimeter \cite{he2006barotropic}, but as explained those values are taken from CMEMS products.

For the rest of the fields, that is baroclinic velocities, salinity, and temperature, which we refer to generically as $\phi$,  ROMS has various types of boundary conditions including open, closed, and periodic. See \cite{Marchesiello2001} for a more thorough exploration of the options. We examined the effects of several of those conditions: mixed radiation-nudging open boundary conditions, clamped boundary conditions, and the use of a sponge layer.

\textbf{Clamped boundary condition.} This is a Dirichlet type of boundary condition. It  consists in setting the boundary value to a known exterior value:
\begin{equation}
    \phi = \phi_{\textrm{ext}}.
\end{equation}
Here $\phi_{\textrm{ext}}$  represents the exterior value of the field $\phi$, provided in our case by the CMEMS  Med-Physics data.

\textbf{Mixed radiation-nudging boundary condition}. 
Open boundary conditions in coastal ocean modeling can be challenging to implement due to the possibility of incoming and outgoing flow at the same boundary or at different depths at the same horizontal location. A mixed boundary condition combining radiation and nudging, first introduced in \cite{Marchesiello2001}, has been proposed as a solution. This boundary condition uses the radiation condition to determine whether the boundary is passive (outward propagation) or active (inward propagation). In the passive case, the solution is allowed to pass through the boundary using radiation extrapolation to prevent excessive reflection. In the active case, the solution is nudged towards external data sources, such as observed climatologies or large-domain model outputs, to provide the necessary information for the dynamical equations without over-specifying the solution. One-way nesting can also be used when the external data is taken from time-evolving large-domain model solutions.

The radiation condition for a prognostic model variable $\phi$ is:
\begin{equation}
    \frac{\partial \phi}{\partial t} = -\left(c_x \frac{\partial \phi}{\partial t} + c_y \frac{\partial \phi}{\partial t} \right), 
\end{equation}
where $(x,y)$ are respectively, the normal and tangential directions to the boundary in local Cartesian coordinates. The phase speeds $(c_x, c_y)$ are projections of the oblique radiation, calculated from the $\phi$ field surrounding the boundary point as follows:
\begin{equation}
    c_x = -\frac{\partial \phi}{\partial t} \frac{\partial \phi /  \partial x}{\left(\frac{\partial \phi}{\partial x}\right)^2 + \left(\frac{\partial \phi}{\partial y}\right)^2},
\end{equation}
and
\begin{equation}
    c_y  = -\frac{\partial \phi}{\partial t} \frac{\partial \phi /  \partial y}{\left(\frac{\partial \phi}{\partial x}\right)^2 + \left(\frac{\partial \phi}{\partial y}\right)^2}.
\end{equation}

In the active boundary, we apply the nudge equation. The nudging layer is a region where model data is relaxed towards external data. For this region, a nudging term is added to the equations of tracers and surface elevation. Its mathematical form is added to the right-hand side of the prognostic equations as follows:
\begin{equation}
    \frac{\partial \phi}{\partial t} = \textrm{r.h.s} - \frac{1}{\tau_{\textrm{nud}}} (\phi - \phi^{\textrm{ext}}).
\end{equation}
In our work, when mixed radiation-nudging boundary condition is used, $\tau_{\textrm{nud}} = 10$ days. The selected value was determined based on a typical advective timescale: Given our grid size of 80 m and maximum velocities of 0.01 m/s, we derive a timescale of 80/0.01 = 8000 seconds, which equates to approximately 10 days for the inflow.

\textbf{Sponge layer.} It is typically used to dump noisy effects at the boundaries  due to discrepancies between the evolving model solution in the domain and the external data in which it is nested. This is implemented by increasing there the coefficients that enhance the horizontal mixing. Both a factor that multiplies the horizontal mixing and its width are indicated 
(see Table \ref{tab:experiments}).

\begin{table}[]
\begin{tabular}{|l|l|l|}
\hline
\multicolumn{1}{|l|}{Parameter} & \multicolumn{1}{l|}{Value} & \multicolumn{1}{l|}{Description}           \\ \hline
$Lm$                              & 344                          & number of points in longitude direction     \\ \hline
$Mm$                              & 228                          & number of points in latitude direction    \\ \hline
$N$                              & 10                          & number of vertical (sigma) levels          \\ \hline
$h_\textrm{max}$                & 146                          & maximum depth of the domain (metres)               \\ \hline
$h_\textrm{min}$                & 0.29                         & minimum depth of the domain (metres)               \\ \hline
$\theta_s$                      & 5.0                          & sigma coordinate surface stretching factor \\ \hline
$\theta_b$                      & 0.4                          & sigma coordinate bottom stretching factor  \\ \hline
$\Delta_t$                      & 10                           & baroclinic time-step (seconds)                       \\ \hline
$\Delta_{tt}$                   & 7                         & barotropic time-step  (seconds)       \\ \hline       
\end{tabular}
\caption{Numerical values for the parameters associated with the setup of the ocean model.}
\label{table:L0}
\end{table}

\section{Remote sensing}
\label{rs}
The waters surrounding the Rafina Port and its vicinity are oligotrophic-like (clear and transparent), which limited the number of available turbidity images that could contribute to the qualification of the Rafina Port hydrodynamic model. To qualify the hydrodynamic model,  data from the Sentinel 2A and 2B Multi-Spectral Instrument (MSI)  were examined for the period ranging from November 1st, 2019 to January 31st, 2020.

Sentinel 2A and 2B had a revisiting cycle of 5 days. Level 1C Top of Atmosphere reflectance products from Sentinel 2 were downloaded from the  \href{https://scihub.copernicus.eu/}{Sentinel Data Hub}.
The standard  \href{http://step.esa.int/main/third-party-plugins-2/sen2cor/}{SEN2COR} library of atmospheric correction algorithms, provided by Scihub Copernicus, for the Sentinel 2 MSI sensor was initially designed for land applications. However, for the purpose of reporting ocean color-derived products and qualifying the hydrodynamic model outputs, Level 2 data required an atmospheric correction algorithm specifically designed to handle the radiometric constraints of water pixels data. As a result, the \href{https://odnature.naturalsciences.be/remsem/software-and-data/acolite}{ACOLITE} toolbox \cite{VanRud} was selected. ACOLITE enabled easy and efficient processing to correct Sentinel-2 Level 1C data and generate value-added derived products for both coastal and inland waters through the use of a dark spectrum algorithm for atmospheric correction \cite{vanhellemont2019}.

\begin{figure}
	\centering
 \includegraphics[width=1\linewidth]{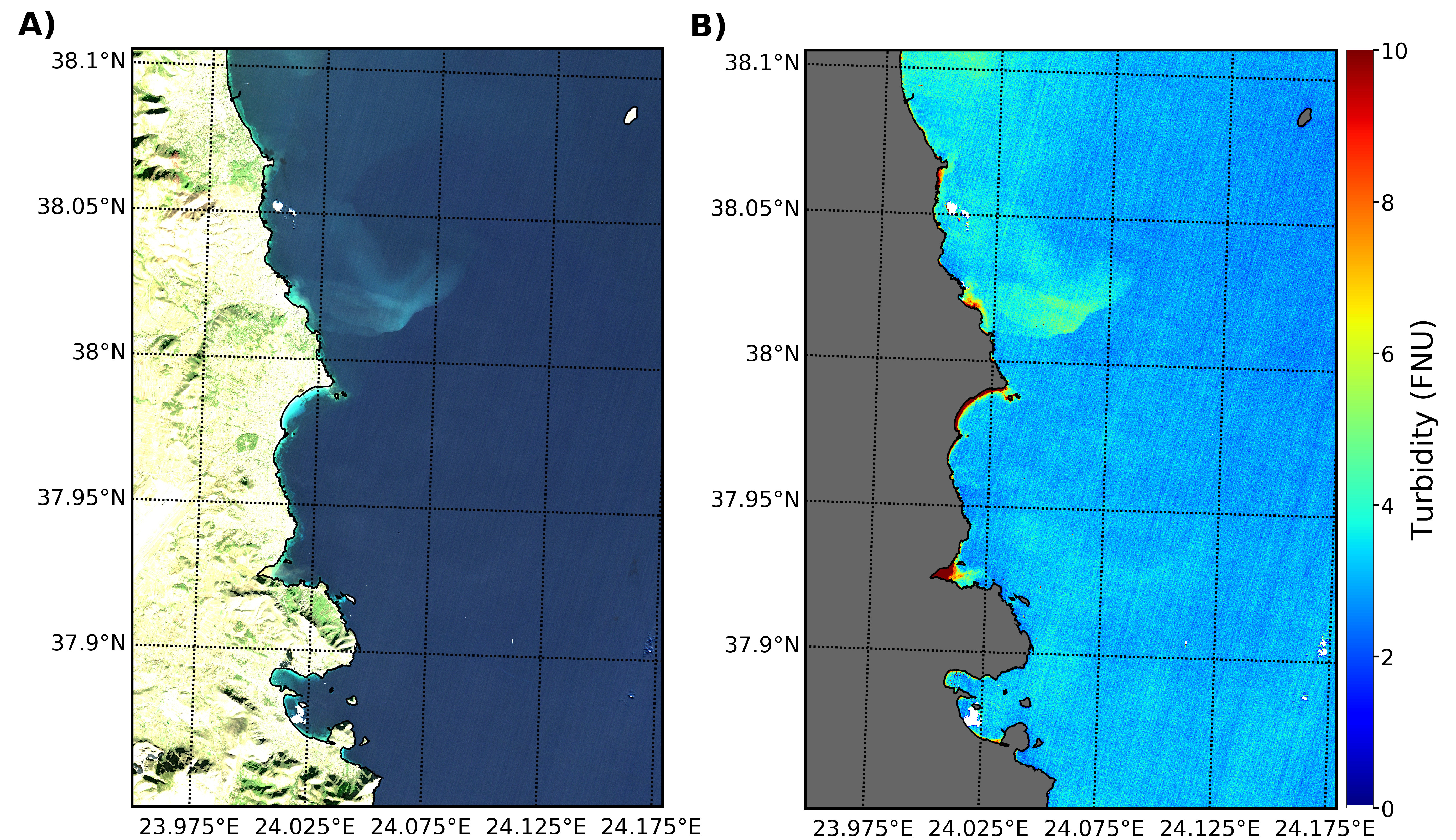}	
	\caption{Sentinel 2 images on the 22nd of November 2019 in the Rafina Port area showing convoluted shapes. (a) Quasi-true RGB color image;(b) turbidity image obtained from ACOLITE algorithms.}
\label{fig:sentinel2}
\end{figure}

The ACOLITE atmospheric correction generates Level 2 images that include biogeochemical-derived products such as turbidity \cite{Nec1} and suspended matter \cite{Nec2}, among others. Additionally, the ACOLITE toolbox includes sunglint correction developed by \cite{har}. These images are useful for model qualification as ACOLITE produces both well-corrected and water discoloration products, including phytoplankton blooms, turbidity, and suspended matter. The spread of these compounds, as evidenced by water discoloration, makes visible the convoluted shapes of the  dynamical objects that we will explain in detail in the next section. As a result, the patterns shown in remote sensing images can be correlated with patterns obtained from the hydrodynamic model. Conversely, in the absence of these compounds, the use of very high-resolution images for this purpose is limited. Figure \ref{fig:sentinel2} shows patterns visible on the sea surface in the Rafina Port area on November 22, 2019, as captured by Sentinel 2 satellite images. Figure \ref{fig:sentinel2}a) depicts a quasi-true color image in RGB (using band 4, band 3, and band 2 of the MSI sensor), while Figure \ref{fig:sentinel2}b) shows a pseudocolor turbidity image in Formazin Nephelometric Units (FNU). 

\begin{figure}
	\centering
	\includegraphics[width=1\linewidth]{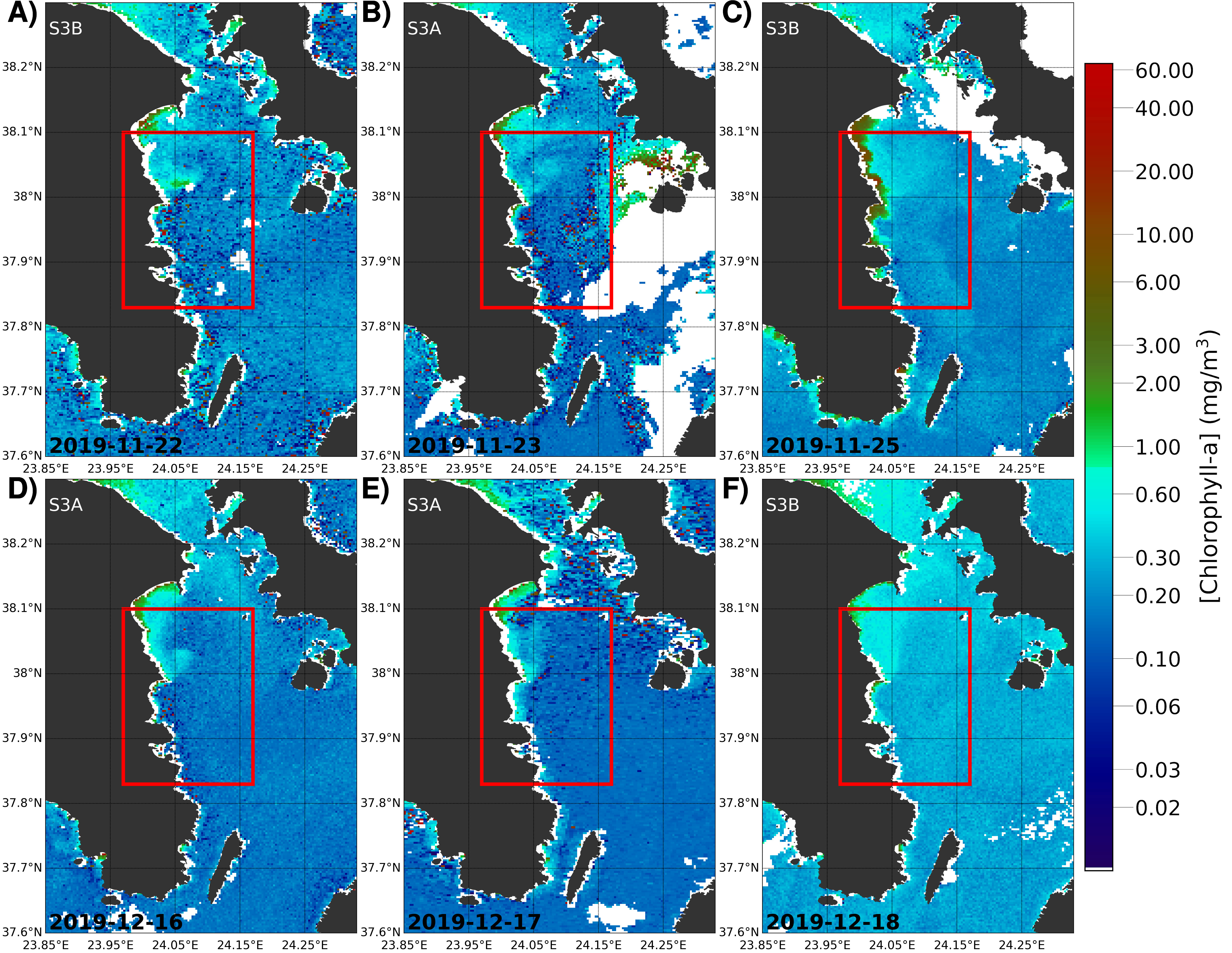}
	\caption{Sentinel 3A and B images in the Rafina Port area showing convoluted shapes of chlorophyll. (a) on the 22th of November 2019, (b) on the 23th of November 2019, (c) on the 25th of November, (d) on the 16th of December, (e) on the 17th of December, and (f) on the 18th of December. Red rectangles mark the region where very high-resolution simulations have been performed.}
\label{fig:sentinel3}
\end{figure}

Ocean and Land Colour Instrument (OLCI) Level 1B data, covering the study area, were acquired from Sentinel 3A and 3B via the \href{https://scihub.copernicus.eu/}{Sentinel Data Hub}. These data underwent processing using \href{https://seadas.gsfc.nasa.gov/}{SeaDAS version 8.2} to produce the required outputs. Ancillary files needed for atmospheric correction were obtained using the "getanc.py" routine. The l2gen SeaDAS binary was utilized to derive geophysical variables, resulting in Level 2 files. These files include remote sensing reflectances and [Chlorophyll-a], among other ocean color products. To produce projected images onto a geographical Lat/Lon WGS84 GeoTIFF, Sentinel 3 Level 2 files were mosaicked by satellite and day using the Graphic Processing Tool (gpt.sh) Mosaic operator. Figure \ref{fig:sentinel3} displays the sea surface patterns captured by Sentinel 3 A and B in the same area. Despite the lower resolution, higher frequency passes enabled the identification of a longer series during the study period. This series also includes an image taken on November 22nd, where Sentinel 2 provided a higher resolution.

\section{Transport and  dynamical systems}
\label{ds}

Images displayed in section \ref{rs} show patterns 
that correspond to  Particulate or Dissolved, Organic or Inorganic Matter (POM, DOM, PIM, DIM),
Chlorophyll, Turbidity, etc., with different origins (primary production blooms, sediments, wastewater plumes).
These light passive particles, suspended in the ocean, are described by a concentration field, $C$, whose general evolution is given by \nicob{Eq.} \eqref{treq}. Assuming that at the surface motion  is mainly horizontal,
\nicob{Eq.} \eqref{treq} can be reduced to the following  advection-diffusion equation: 
\begin{equation}
    \frac{\partial C}{\partial t} + \textbf{v}_h\cdot \nabla C =\mu \nabla^2C,
    \label{advection-difussion-eq}
\end{equation}
where $\mu$ is the horizontal molecular diffusion and $\textbf{v}_h$ is the vector with  the horizontal velocity components of the ocean model. Assuming that at our scales advection is the dominant factor in transport   (see    \cite{GGS2021, GGS2022, Lekien2005, Olascoaga2012, mh370}) then $\mu \sim 0$
 and \nicob{Eq.} (\ref{advection-difussion-eq}) reduces to:
\begin{equation}
    \frac{\partial C}{\partial t} + \textbf{v}_h\cdot \nabla C = \frac{d C}{dt} = 0.
    \label{advection-eq}
\end{equation}
This equation assumes that $C$ is conserved along the fluid parcel trajectories. 
Fluid parcels on the ocean surface follow trajectories   in longitude/latitude coordinates $\vec{x}(t)=(\lambda(t), \phi(t))$ 
that evolve according to the dynamical system:
\begin{eqnarray}
  \frac{d\lambda}{dt} &=& \frac{u(\lambda,\phi,t)}{R \cos\phi}, \nonumber \\
  \frac{d\phi}{dt} &=& \frac{v(\lambda,\phi,t)}{R}.\label{eq-motion}
\end{eqnarray}
where $R$ is the Earth's radius and $u$ and $v$ are the eastwards and  northwards velocity components  provided by the hydrodynamical model at the surface layer.  

In order to characterize the performance of  $u$ and $v$ 
to consistently describe patterns observed in satellite  images in figures \ref{fig:sentinel2} and \ref{fig:sentinel3}, we are 
going to extract geometrical structures from Eq. \eqref{eq-motion} using 
specific velocity fields obtained under different settings of the hydrodynamical model. 
The geometrical structures that we will use,  the Lagrangian Coherent Structures (LCS), provide a signature for a specific velocity field that highlights the essential transport features associated with it.  
LCS form time-dependent material surfaces that separate fluid paths that behave differently. An important mathematical result is that these Lagrangian patterns are robust against velocity field perturbations, whereas the comparison of two individual trajectories is not. This allows a better characterization of the velocity fields produced by the model, by allowing to establish as different outputs those that have a different Lagrangian signature, while those that present a similar Lagrangian pattern, can be characterized as similar.
\nico{This approach has found successful application in the context of pollution \cite{GGS2022b}. }

\nico{ We compute the LCS  with the method known as Lagrangian Descriptors (LDs).} The particular LD that we use is a function referred to as $M$ \cite{mancho2013, madrid2009} which is defined as follows:
\begin{equation}
    M(\vec{x}_0,t_0, \tau) = \int_{t_0-\tau}^{t_0+\tau} \|\textbf{v}_h(\vec{x}(t),t)\| = \int_{t_0}^{t_0+\tau}\|\textbf{v}_h(\vec{x}(t),t)\| + \int_{t_0-\tau}^{t_0} \|\textbf{v}_h(\vec{x}(t),t)\|.
    \label{ld}
\end{equation}
where $\|\cdot\|$ stands for the modulus of the velocity vector. In this expression, the computation of the function $M$ is split into its forward time ($M^f$) and backward time ($M^b$) contributions. At a given time $t_0$, the function $M(\vec{x}_0,t_0, \tau)$ measures the arc length traced by the trajectory starting at $\vec{x}_0 = \vec{x}(t_0)$ as it evolves forward and backward in time for a time interval $\tau$. For a sufficiently large $\tau$, 
the function $M$ exhibits a distinct structure that reveals singular features emphasizing the Lagrangian Coherent Structures (LCS). Figure \ref{fig:quantcomp} displays the calculation of $M^b$ in the vicinity of Rafina port for $\tau=3$ days on selected dates consistent with those of figure \ref{fig:sentinel3}.
\nico{To compute figure \ref{fig:quantcomp}, we perform trajectory calculations from Eq. \eqref{eq-motion} by integrating them using a 5th-order Runge-Kutta method. The arc length \nicob{expressed in Eq. \eqref{ld}, } is then determined by summing the lengths of linear segments connecting consecutive steps in the Runge-Kutta method.}

In \nicob{Eq.} \eqref{ld} the backward integration ($M^b$) highlights the structure of attracting LCS and we expect that suspended material will eventually, after a transient time, be aligned with these features. Patterns 
visible from satellite imagery
correspond to suspended material 
that could  decay towards these features \cite{GGS2021,GGS2022,GGS2022b}. Additionally
dynamical barriers on the sea surface could also be made visible by the suspended matter, as barriers  would keep the suspended material on both sides  unmixed as far as the original distribution of the suspended material was only on one side of the barrier.


\begin{table}[]
\begin{tabular}{|l|l|l|}
\hline
\multicolumn{1}{|l|}{Parameter} & \multicolumn{1}{l|}{Value}          \\ \hline
Horizontal Viscosity $A_{u,v}$& 2 $m^2/s$\\ 
\hline
Horizontal Diffusivity $A_{T,S}$& 2 $m^2/s$\\ 
\hline
Bottom Drag factor $C_d^b$ & 0.003 \\ 
\hline
Sponge factor & 1 \\ 
\hline
Wind drag factor $C_d^s$ & $6.0 \cdot 10^{-5}$ \\ 
\hline
Bathymetry & NAVIONICS \\
\hline
Boundary conditions & Clamped \\ 
\hline
\end{tabular} 
\caption{This table presents the numerical values associated with each parameter utilized in the base solution}
\label{tab:bs}
\end{table}
\section{Numerical experiment setups and performance quantification }
\label{nexppq}

The hydrodynamical model can be set up with different parameters and boundary conditions, which can lead to different outputs, particularly in the currents. Typically, these free parameters are adjusted by benchmarking ocean model outputs against in situ Eulerian metrics obtained from acoustic Doppler current profilers (ADCPs), mooring data, and other sources. In this work, we aim to evaluate the performance of these models from a Lagrangian perspective by using satellite imagery, as described in the previous section.

We begin by establishing a configuration for a base solution (BS) with parameters provided in Table \ref{tab:bs}, which complement those shown in Table \ref{table:L0} and are kept constant.
We execute this configuration during the timeframe for which reference images are accessible. These consist of two consecutive periods in 2019 (refer to Figures \ref{fig:sentinel2} and \ref{fig:sentinel3}), with a total of six images. The first period ranges from 2019-11-22 to 2019-11-25, while the second period spans from 2019-12-15 to 2019-12-18.
To ensure that these outputs are not dependent on the initial conditions used, the system is initialized prior to the period of interest and run for a sufficient period of time to converge to the global attractor. Given that our system is subject to time-dependent boundary conditions (such as winds and open lateral boundary conditions), the type of attractor involved is called a pullback attractor \cite{medjo}. When LCS are computed using velocity fields from different initial conditions, there are no discrepancies observed once the solutions have converged to the pullback attractor.

\begin{figure}[h]
    \centering
    \includegraphics[width=1\linewidth]{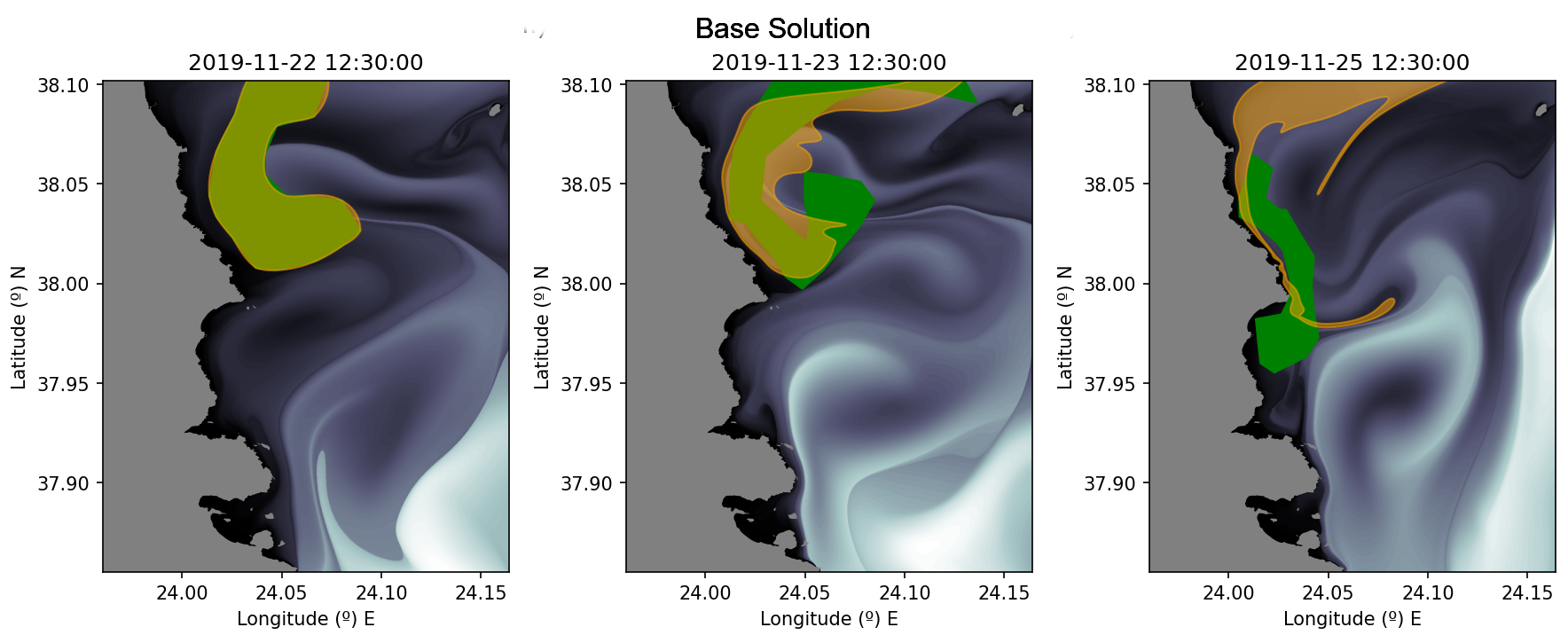}\\
    \includegraphics[width=1\linewidth]{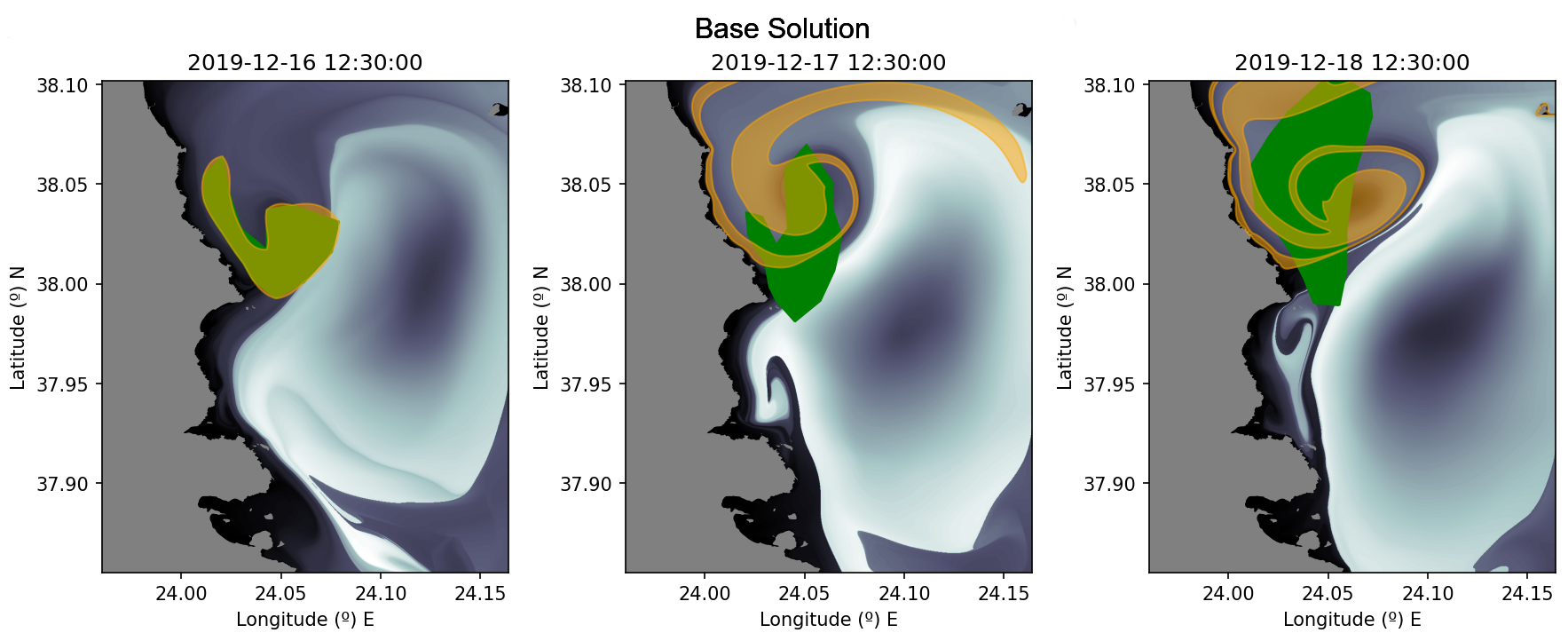}
    \caption{Evaluation of $M^b$ on the BS model for $\tau=3$ days. Patterns are displayed in a gray tone. Panels a), b), and c) show the results on the time series correlated to the satellite imagery displayed in panels a), b), and c) of figure \ref{fig:sentinel3}. Panels   d), e), and f) show the results on the time series correlated to the satellite imagery displayed in panels d), e), and f) of the same figure. The green blobs show the contours of an estimation of the evolution of chlorophyll from the satellite images. The brown blobs show the evolution according to the model of the green blob at the initial date of each time series. }
    \label{fig:quantcomp}
\end{figure}

Defining a metric that provides a quantification to compare
the sequence of satellite observations  with the performance of the hydrodynamic model outputs is not an easy task. 
The reason for this difficulty is that it involves comparing complicated and subtle dynamical objects.
To accomplish this, we have proposed a methodology that makes assumptions about both the observations and the models. 
Regarding observations, we have extracted polygons from the satellite imagery that represent zones where the chlorophyll is homogeneously distributed.
Figure \ref{fig:quantcomp} shows a sequence of green polygons  for the two periods. Assuming that these structures are formed by suspended particles and  are  purely advected by the currents, they provide a ground truth or benchmark to be recovered by the model. 
This figure also shows, in brown, the evolution of the chlorophyll blobs that were initialized on the first day of the sequence by using the velocity fields from the base solution set. \nicob{This calculation has been implemented using ideas by \cite{Mancho2003}.} In the background, for interpretation purposes, we have represented the evaluation of $M^b$  for $\tau=3$ days, which highlights, through singular features,  the attracting material curves of the flow.  There is a nice consistency between these features and the evolution of the brown blob.
Ideally, if the model is good we would expect a lot of overlap between the green and brown blobs.
Certainly, for the suspended particulate matter there may be sinks and/or sources and this is reflected in the fact that the sequence of green blobs does not preserve the area. Additionally, diffusion adds to spreading on the ocean surface. These phenomena are not described by pure advection, but still, the overlap is an intuitive measure to consider.
  For each sequence, in  the days after  the first one, we establish as a measure the area fraction of  the green blob that  is covered with the brown blob.  This will be a number $\cal{A}$ satisfying $0\leq {\cal A}\leq 1$ that we interpret as follows, close to 1 means good performance while close to 0 implies poor performance. We will comment further on these images in the results section.

We complete this section by describing a series of free parameter configurations in the hydrodynamic model. These are defined from variations of the parameters taken for the base solution and define a series of experiments. Our study aims to qualify the results of all these experiments, using the newly defined metric.
The experiments are divided into six categories:
 
\begin{itemize}
\item C1: Varying horizontal mixing coefficients (viscosity and diffusivity) in two experiments compared to the Base Solution (BS).
\item C2: Investigating two different bottom drag coefficients in comparison to the BS.
\item C3: Assessing two distinct surface forcing wind stress fields relative to the BS.
\item C4: Conducting a single experiment with bathymetry replaced by GEBCO data.
\item C5: Implementing a sponge layer with six times the base horizontal viscosity and diffusivity in one experiment.
\item C6: Executing two experiments where the open boundary condition (OBC) was replaced by the mixed radiation-nudging boundary condition, modifying the factor between passive (outflow) and active (inflow).
\end{itemize}

Details of the parameters taken for each   experiment can be found in Table \ref{tab:experiments}.


\begin{table}[h]
\centering
{Summary of experiments}\\
\begin{tabular}{|l|l|l|}
\hline
\textbf{Category} & \textbf{Experiment} & \textbf{Parameters} \\
\hline
C1: Horizontal Mixing Coefficient & Exp 1 & $A_{u,v} = A_{T,S} = 0.2 m^2/s$ \\
\cline{2-3}
& Exp 2 & $A_{u,v} = A_{T,S} = 5 m^2/s$ \\
\hline
C2: Bottom Drag Coefficient & Exp 1 & $C_d^b = 0.03$ (non-dimensional) \\
\cline{2-3}
& Exp 2 &\nico{ $C_d^b = 3\cdot 10^{-5}$} (non-dimensional) \\
\hline
C3:  Wind stress & Exp 1 & $C_d^s = 6\cdot 10^{-4}$ \\
\cline{2-3}
& Exp 2 & $C_d^s = 6\cdot 10^{-7}$ \\
\hline
C4: Modified Bathymetry & Exp 1 & GEBCO data (Fig. \ref{fig:rafina} b)) \\
\hline
C5: Sponge at Boundary & Exp 1 & $6\cdot A_{u,v,T,S}$ \\
\hline
C6: Mixed Radiation-Nudging BC & Exp 1 & No sponge, $\tau_{\textrm{nud}} = 10$ days,  \\
\cline{2-3}
& Exp 2 & Sponge, $\tau_{\textrm{nud}} = 10$ days \\
\hline
\end{tabular}
\caption{Setting the varying parameters for defining categories and experiments to conduct sensitivity tests.}
\label{tab:experiments}
\end{table}

\section{Results}

\label{res}
First, we comment on the results displayed in Figure \ref{fig:quantcomp}. The upper row shows a sequence of satellite imagery captured between November 22nd and 25th, 2019, where chlorophyll is visible. On November 23rd, the tracer appears to swirl, while on November 25th, the chlorophyll in the upper half of the domain has diffused, weakening all features except for a large concentration aligned with the coast. A material barrier transverse to the coast at around 38 degrees North appears to persist in all these upper panels.
The dynamical structure of the currents of the BS model is obtained from $M^b$, whose pattern is displayed in the gray tones in the background. The attracting material curves are visible through the singular features and consistently show the transverse barrier to the coast  present in  the observations. 
  The evolution of the blob according to the BS model is depicted in brown, which also exhibits swirling (panel b) and contraction towards the coast (panel c).
The bottom row displays satellite imagery where chlorophyll is visible between December 16th and 18th, 2019. On December 16th, the chlorophyll tracer has a swirling shape. The boundary of this shape progressively weakens on December 17th and 18th. A pronounced transverse barrier to the coast in a northeasterly direction evolves progressively, bringing its alignment closer to the north.
The $M^b$ pattern  is shown in the gray tones in the background of the image. It confirms the presence of a barrier   in a northeasterly direction, although with a slightly different inclination than that of the satellite images.
According to the BS model,  the brown blobs swirl counterclockwise as they evolve.  The first row of Table \ref{tab:relative_intersections} provides values of the computed ${\cal A}$ for the first chlorophyll series on November 23rd and 25th, 2019, and for the second chlorophyll series on December 17th and 18th, 2019, quantifying the performance of the BS model. In addition to this data, Table \ref{tab:experiments_summary} shows the averages of these quantities for the first and second periods as well as for the total period.

\begin{figure}[h]
    \centering
    \includegraphics[width=1\linewidth]{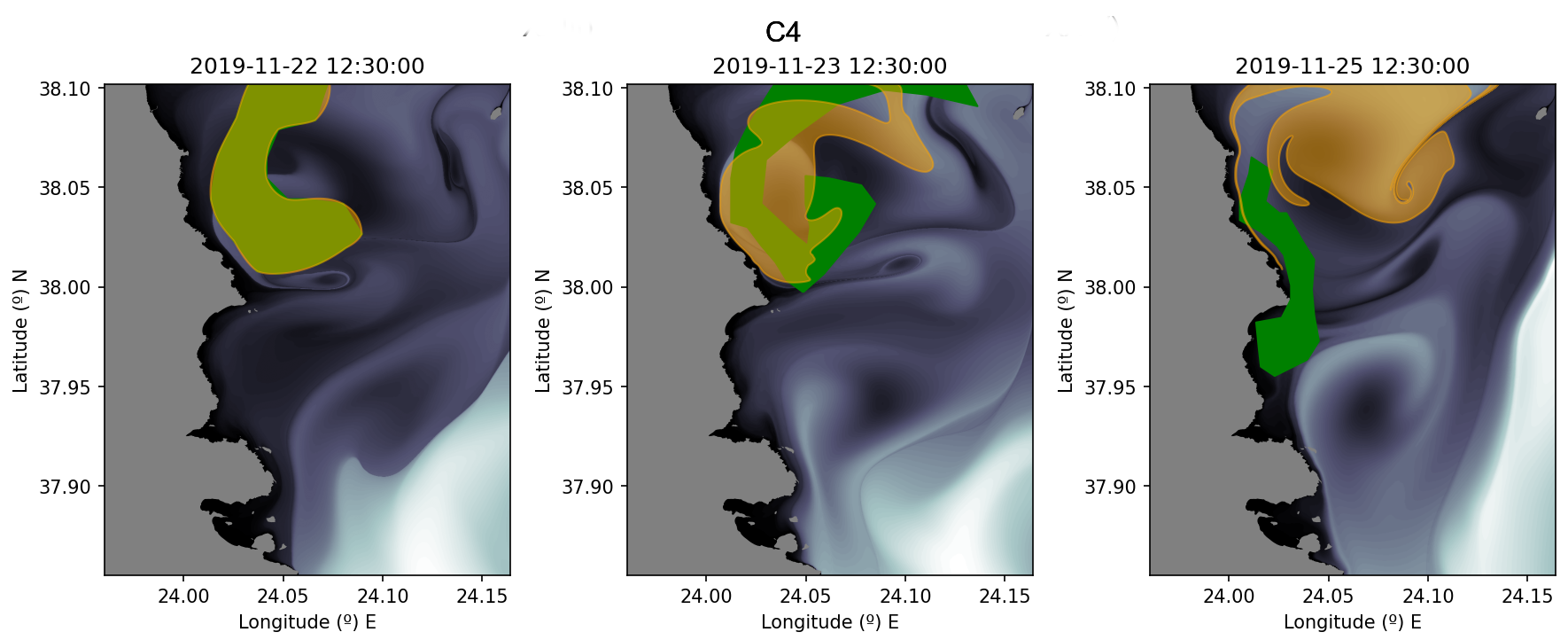}\\
    \includegraphics[width=1\linewidth]{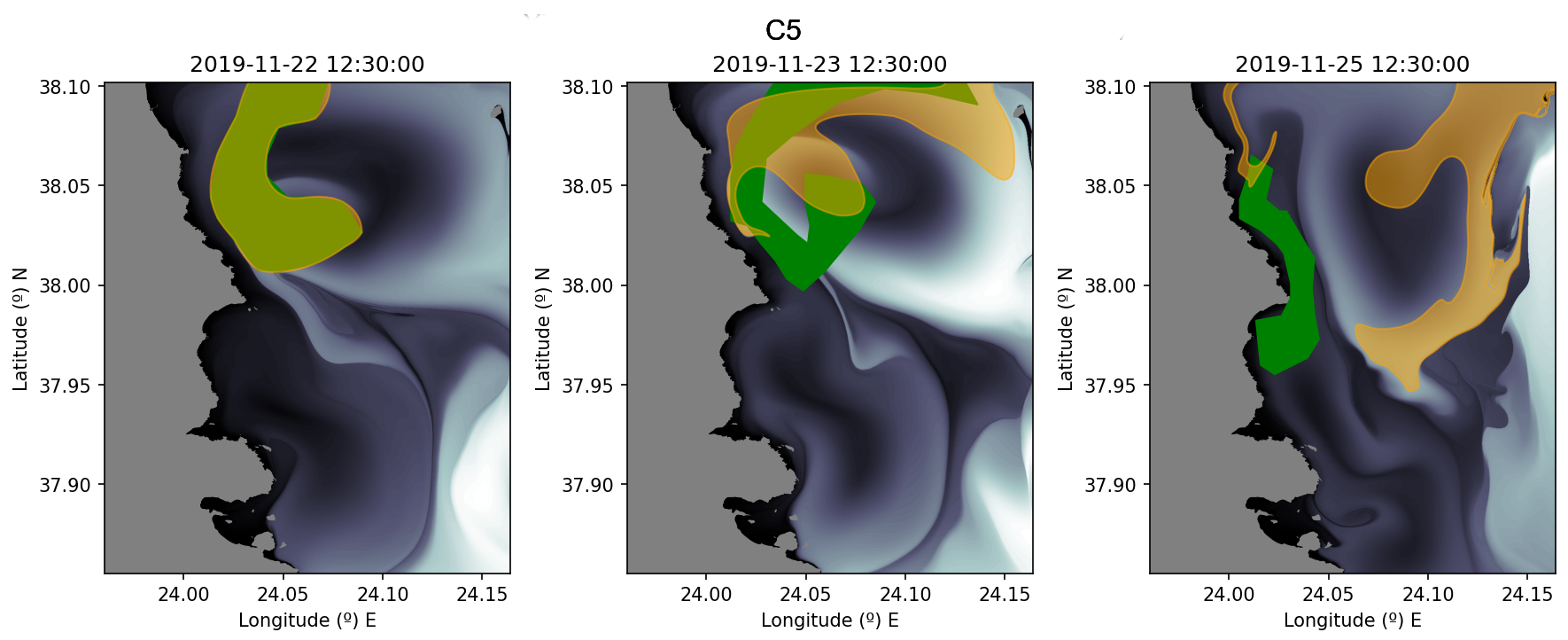}\\
    \includegraphics[width=1\linewidth]{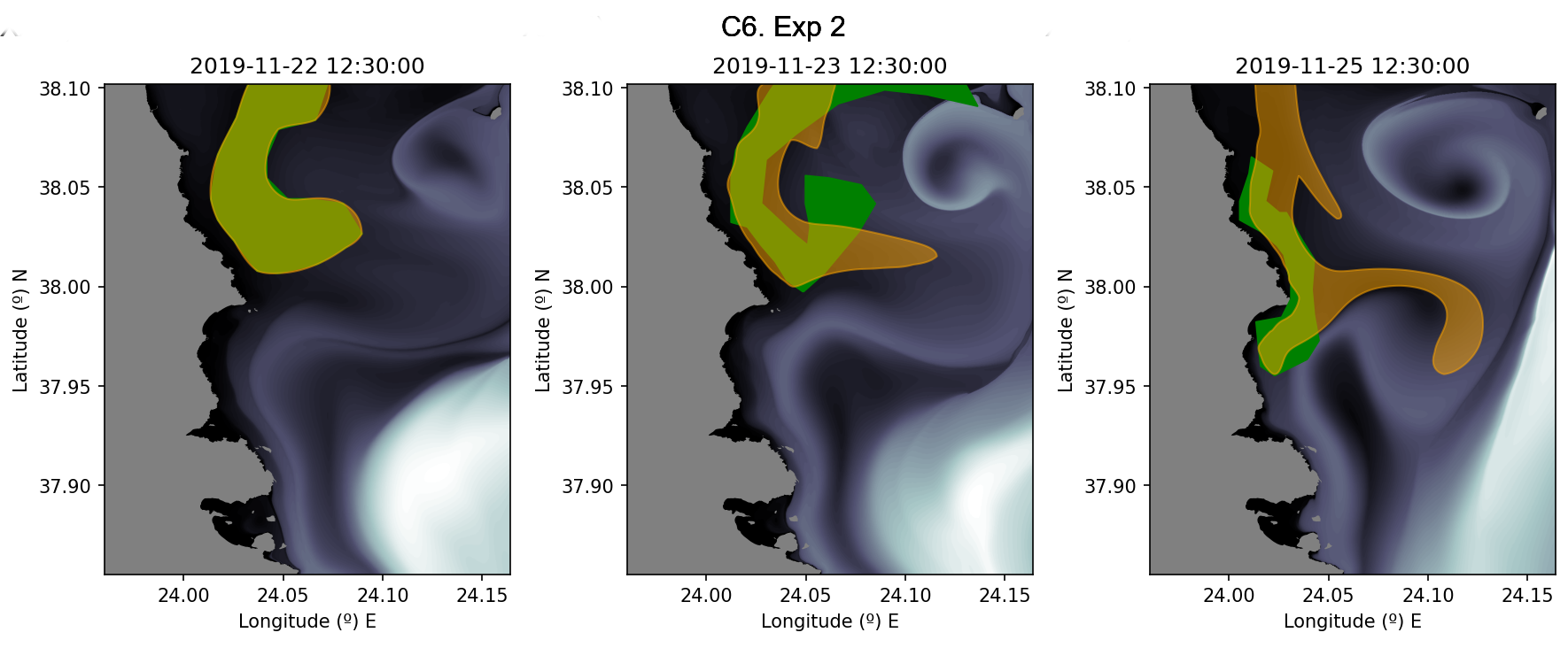}
    \caption{Evaluation of $M^b$ on several models for $\tau=3$ days. Patterns are displayed in a gray tone. The first, second, and third columns  show the results on the time series correlated to the satellite imagery displayed in panels a), b), and c) of figure \ref{fig:sentinel3}. The first row shows the results for Category 2,  experiment 2, on the bottom drag condition.  The second row shows the results for Category 5,  experiment 1, on the sponge conditions. The third row shows the results for Category 6,  experiment 2, on the mixed radiation-nudging conditions. The green blobs show the contours of an estimation of
the evolution of chlorophyll from satellite images. The brown blobs show the evolution according to each
model of the green blob at the initial date of each time series. }
    \label{fig:quantcomp1}
\end{figure}

Tables \ref{tab:relative_intersections} and \ref{tab:experiments_summary} also provide quantification of the performance of all experiments. Results for ${\cal A}$ fluctuate between 0 and 0.75 for individual experiments, and between 0.2 and 0.65 for averaged values. Averaged values of ${\cal A}\sim 0.2$ or individual values of ${\cal A}\sim 0.1$ indicate poor performance.
During the first period, extremely low performances on individual days are obtained for instance  on 25, November 2019 
for the category 4  model that uses the GEBCO bathymetry (C4, exp1) 
and for the sponge layer experiment (C5,exp1) that introduces high diffusivity and viscosity on the boundaries of the domain.
Explicit representations for selected poor-performing cases are provided in  
Figure \ref{fig:quantcomp1}, where the top two rows are represented the cases (C4,exp1) and 
(C5,exp1).
The highest-performing model for the first period corresponds to  the (C6, exp2) experiment that implements a mixed sponge and radiation-nudging condition. The last row in figure \ref{fig:quantcomp1} provides a visualization of the performance of this model. Remarkably, its dynamical structure allows the blob to be maintained elongated along the coast, although the model also shows incursions towards the interior of the sea.

\begin{figure}[h]
    \centering
    \includegraphics[width=1\linewidth]{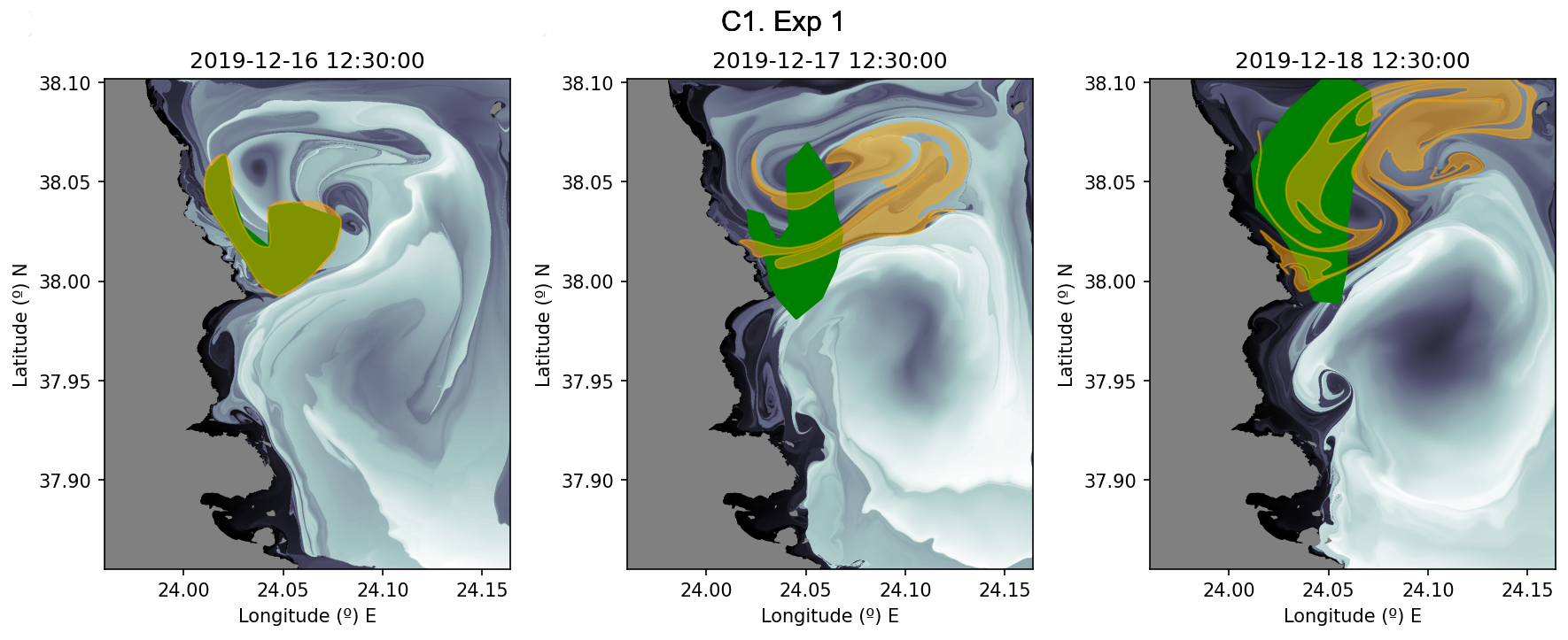}\\
    \includegraphics[width=1\linewidth]{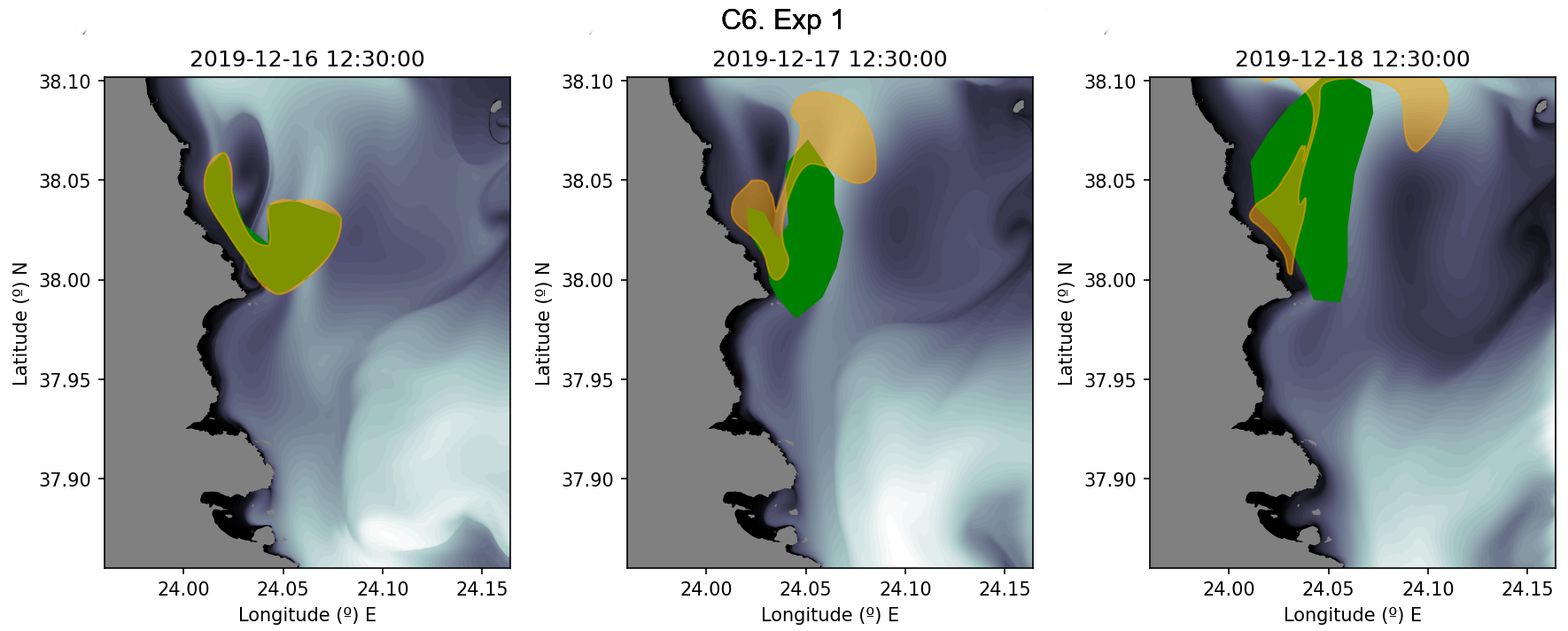}\\
    \includegraphics[width=1\linewidth]{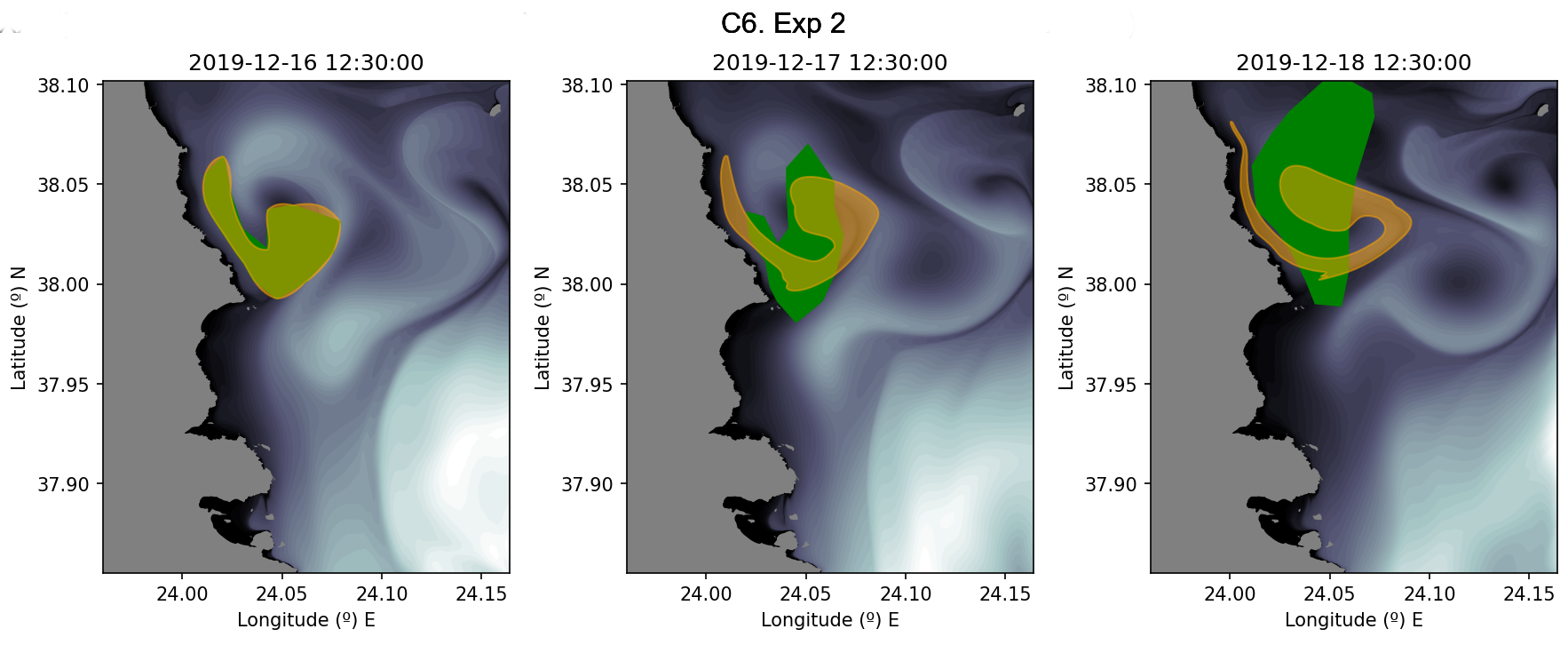}
    \caption{Evaluation of $M^b$ on several models for $\tau=3$ days. Patterns are displayed in a gray tone. The first, second, and third columns  show the results on the time series correlated to the satellite imagery displayed in panels d), e), and f) of figure \ref{fig:sentinel3}. The first row shows the results for Category 1,  experiment 1, on the horizontal diffusion and viscosity coefficients.  The second row shows the results for Category 3,  experiment 1, on the wind stress conditions. The third row shows the results for Category 6,  experiment 2, on the mixed radiation-nudging conditions. The green blobs show the contours of an estimation of
the evolution of chlorophyll from satellite images. The brown blobs show the evolution according to each
model of the green blob at the initial date of each time series.}
    \label{fig:quantcomp2}
\end{figure}

Regarding the second period, the top two rows of Figure \ref{fig:quantcomp2} represent simulations performed with the (C1, exp1) model, which decreased the horizontal viscosity and diffusivity coefficients, and the (C6, exp1) model, which considers only a radiation-nudging condition. Both simulations perform poorly. The (C1, exp1) model presents invariant dynamical structures that are very rich and involve very small scales. Some of these structures remain within the observed green blob. The (C6, exp1) model performs poorly. The absence of a sponge effect at the boundary in the (C6, exp1) model influences the solution compared to what is obtained by considering it in the (C6, exp2) model.
The latter is illustrated in the bottom row, the (C6, exp2) model. According to the values given in Table \ref{tab:relative_intersections}, the performance is rather good on the 17th but decays on December 18th, 2019.

The effects of different models performing differently on different days and periods can be clearly observed in Table \ref{tab:relative_intersections} and Table \ref{tab:experiments_summary}. For example, model (C6, exp2) obtained with the mixed sponge radiation-nudging condition
 performs rather consistently in the first period, but worse in the second period. The same occurs for the experiment that considers only the radiation-nudging condition (C6, exp1). On the contrary, the model (C1, exp2) obtained by increasing horizontal viscosity and diffusivity performs more consistently in the second period, than in the first period.
 Similarly occurs for the model (C2, exp1) which increases the bottom drag coefficient. 
All in all, the total average column in Table \ref{tab:experiments_summary} confirms that (C6, exp2), with the mixed radiation-nudge condition, 
has the best performance and the initial BS choice has moderate performance.  

\begin{table}[]
\scalebox{1}{
\begin{tabular}{|l|l|l|l|l|}
\hline
\textbf{Experiments} & \textbf{2019-11-23} & \textbf{2019-11-25} & \textbf{2019-12-17} & \textbf{2019-12-18} \\ \hline
B.S & 0.49919 & 0.085 & 0.3669 & 0.2720 \\
\hline
C1. Horizontal mixing  exp 1 & 0.3953  & 0.1185 & 0.2065 & 0.2802 \\
\hline
C1. Horizontal mixing  exp 2 & 0.4654 & 0.3188 & 0.4875 & 0.3815 \\
\hline
C2. Bottom drag exp 1 & 0.4224 & 0.1096 & 0.3618 & 0.4835 \\
\hline
C2. Bottom drag exp 2 & 0.4375 &  0.3165 & 0.2282 & 0.3676 \\
\hline
C3. Wind stress exp 1 & 0.4462 & 0.4343 & 0.3164 & 0.3224 \\
\hline
C3. Wind stress exp 2 & 0.4883 & 0.087 & 0.3793 & 0.2848 \\
\hline
C4. Modified bathymetry exp 1 & 0.4491 & 0.005& 0.1305 & 0.2363 \\
\hline
C5. Sponge exp 1 & 0.3778 & 0.0278 & 0.29965 & 0.2528 \\
\hline
C6. Mixed radiation-nudging exp 1 & 0.5315 & 0.3682 & 0.2337 & 
0.1660 \\
\hline
C6. Mixed radiation-nudging exp 2 & 0.5555 & 0.7476 & 0.5192 & 0.2551 \\
\hline
\end{tabular}
}
\caption{Results of measured index ${\cal A}$ obtained for the sensitivity experiments. they include the first period (2019-11-23 and 2019-11-25), and the second period (2019-12-17 and 2019-12-18).}
\label{tab:relative_intersections}
\end{table}

\begin{table}[]
\centering
\begin{tabular}{|l|l|l|l|}
\hline
\textbf{Experiments} & \textbf{Avg First Period} & \textbf{Avg Second Period} & \textbf{Total Avg} \\ \hline
B.S. & 0.2922 & 0.3195 & 0.3058 \\
\hline
C1. Horizontal mixing exp 1 & 0.2569 & 0.2433 & 0.2501 \\
\hline
C1. Horizontal mixing exp 2 & 0.3921 & 0.4345 & 0.4133 \\
\hline
C2. Bottom drag exp 1 & 0.2660 & 0.4226 & 0.3443 \\
\hline
C2. Bottom drag exp 2 & 0.3770 & 0.2980 & 0.3375 \\
\hline
C3. Wind stress exp 1 & 0.4402 & 0.3194 & 0.3798 \\
\hline
C3. Wind stress exp 2 & 0.2876 & 0.3321 & 0.3099 \\
\hline
C4. Modified bathymetry exp 1 & 0.2273 & 0.1834 & 0.2054 \\
\hline
C5. Sponge exp 1 & 0.2028 & 0.2762 & 0.2395 \\
\hline
C6. Mixed radiation-nudging exp 1 & 0.4498 & 0.1998 & 0.3249 \\
\hline
C6. Mixed radiation-nudging exp 2 & 0.6515 & 0.3871 & 0.5194 \\
\hline
\end{tabular}
\caption{Averages of  measured index ${\cal A}$ for the first and second periods.}
\label{tab:experiments_summary}
\end{table}


\section{Discussion and Conclusions}
\label{con}

In this paper, we explore a methodology based on the use of dynamical systems ideas to assess the quality of results from different configurations of a high-resolution coastal ocean model.  Our aim is to leverage satellite imagery, which provides observations at a lower cost than in situ observations, and to propose a strategy for quantifying the quality of model results using an ${\cal A}$ index. The models are defined by specifying parameters or boundary conditions of the system with various options that are not known beforehand.

\nicob{The method has used sequences of RGB satellite images that detect suspended chlorophyll in ocean waters.  Despite its proven usefulness, this methodology also has limitations. One is that these image sequences are not always easy to observe. On the one hand, the suspended matter is not always found there, since it depends on the existence of conditions that lead to its proliferation in ocean waters. Furthermore, even if satellites have a reasonable revisit frequency for the area of interest, very often the requested information in the RGB spectral range is obscured by clouds. We have focused on Sentinel 3 images, which have offered us two interesting time series in the period of study. While Sentinel 2 images would have been desirable to increase the resolution of the proposed tests, for instance, down to the very small scale of Rafina Port in the area, we were unable to obtain an appropriate time series for this product.}

\nicob{The computational experiments are implemented in ROMS. ROMS allows easy configuration of the various parameters used in the model by simply changing values in an input file.  However, in the performed experiments, setting different boundary conditions may require additional effort. In particular, utilizing different bathymetries or coupling with lateral and upper boundary conditions involves preparing distinct data files. The selection of the different experiments requires prior work that leads to ruling out other options. The turbulence model implemented could have deserved its own article, to discuss performances and comparisons between all the possibilities available in ROMS \cite{warner2005performance}. However, we have restricted our evaluation by choosing a turbulent model that has been frequently used in similar environments (see for example \cite{SAMOA}).  }

\nicob{ The outlined relationship between satellite images and simulation results is based on certain assumptions about the suspended matter. In particular, we have accepted that chlorophyll behaves like
a purely advected passive scalar. This may be a rough approximation since chlorophyll is found in the phytoplankton, which in turn reacts with nutrients, temperature, and wind. This could have an impact on how the ground truth domains shown in green in Figures  \ref{fig:quantcomp1} and \ref{fig:quantcomp2} are selected. However, as explained below, despite the approximations considered, the results we obtain are consistent with other assessments carried out in similar environments.
}

We have found that many model configurations produce reasonable results and that the choice of boundary conditions for nesting has a significant impact. Among these, radiation-nudging lateral boundary conditions appear to be particularly suitable. \nicob{The advantage found for this open boundary condition is aligned with the upgrade implemented by Spanish Port Authorities in SAMOA models \cite{sotillo2022}, where it has been reported to provide better consistency with the parent solution. On the other hand, our results also confirm the sensitivity of currents to bathymetry, something also reported in \cite{bathy}.} The coupling with the wind at the surface, the adherence at the bottom, and parameters related to the horizontal mixing can all cause distortions in the solutions, leading to deviations from observations.

\nicob{The results presented in this paper highlight the anticipated utility of the proposed methodology for fine-tuning high-resolution coastal models in regions where in situ observations are scarce. Our adjustment is discussed in the framework of Lagrangian transport problems, which have critical applications in search and rescue operations, oil spill responses, and understanding the evolution of harmful algae proliferating due to climate change, impacting fishing, tourist beaches, and more. Lagrangian transport problems are often one of the applications of ocean models most sensitive to inefficient modeling. Furthermore, our methodology illustrates new possibilities for the application and exploitation of satellite images in an era when Earth observation has become one of the largest efforts of global space agencies.}

{\bf Acknowledgements.}
The authors thank E.U. Copernicus Marine Service Information for providing data. 
The authors thank E.U. Copernicus Sentinel Data for providing Sentinel 2 and 3 data used in the present paper. The authors thank REMSEM — OD Nature Remote Sensing and Ecosystem Modelling team for providing acolite software.
AMM and GGS are members of  two CSIC Interdisciplinary Thematic Platforms: POLARCSIC and TELEDETECT.

{\bf Funding.}
The authors acknowledge support from IMPRESSIVE, a project
funded by the European Union’s Horizon 2020 research and
innovation programme under grant agreement No 821922. 
GGS and AMM acknowledge the support of a CSIC PIE project Ref. 202250E001;  the support from grant PID2021-123348OB-I00 funded by  
MCIN/ AEI /10.13039/501100011033/ and by
FEDER A way to make Europe. 

{\bf Competing Interests.}
The authors have  had no competing interests  during the completion of this work.

{\bf Data Availability.}
The datasets generated during and/or analysed during the current study are available from the corresponding author upon reasonable request.

{\bf Declaration of Generative AI and AI-assisted technologies in the writing process.}
During the preparation of this work, AMM used Google Translator, Grammarly, and Chat GPT to improve the use of English in writing. 
After using these tools/services, AMM reviewed and edited the content as necessary and takes full responsibility for the content of the publication.

\bibliographystyle{plain}   
\bibliography{biblio}

\end{document}